\documentclass[12pt,preprint]{aastex}

\newcommand{\Hiirs}{\ion{H}{2}~regions}

\newcommand{\msun}{M$_\odot$}
 
\shorttitle{Chemical Evolution Models}
\shortauthors{Carigi, Peimbert, \& Peimbert}

\received{}
\begin{document}

\title{
The last 5 Gyr of Galactic chemical evolution based on \ion{H}{2} region abundances
derived from a temperature independent method
}

\author {Leticia Carigi\footnotemark[1]}
\email{carigi@astro.unam.mx}

\author{Manuel Peimbert\footnotemark[1]}
\email{peimbert@astro.unam.mx}

\and

\author{Antonio Peimbert\footnotemark[1]}
\email{antonio@astro.unam.mx}

\footnotetext[1]{Instituto  de Astronom\'\i a, Universidad Nacional Aut\'onoma de M\'exico, Apdo. Postal 70-264, M\'exico, C.P. 04510, CdMx, Mexico}

\begin{abstract}
Most of the chemical evolution models are not very reliable for the last 5~Gyr of 
galactic evolution; this is mainly because abundance gradients found in the 
literature show a big dispersion for young objects; a big culprit of this is the 
dispersion found in \ion{H}{2} region gradients. Part of this dispersion arises from two 
different methods used to determine O/H in \ion{H}{2} regions: 
the direct method (DM), based on forbidden lines; and the temperature 
independent method (TIM), based on permitted lines; the differences between these two 
methods are 
about 0.25~dex. We present two chemical 
evolution models of our galaxy to fit the O/H gradients of 
\ion{H}{2} regions, one obtained from the DM and the 
other obtained from the TIM. We find that the model based on the TIM produces an excellent
fit to the observational stellar constraints (B-stars, Cepheids, and the Sun), while 
the model based on the DM fails to reproduce them.
Moreover the TIM model reproduces the flattening observed in the $3-6$ kpc galactocentric range; 
this flattening is attained with an inside-out star formation quenching in the inner disk starting $\sim9$ Gyr ago. 
\end {abstract}


\section{Introduction}\label{intro}

Our knowledge of the Milky Way (MW) is considerably wider and deeper than our knowledge of any 
other galaxy; consequently the MW should be used as a benchmark for the study of
other galaxies. Therefore, it is paramount to have a reliable and robust model of the Galaxy
to improve the development of future models of other galaxies; a key ingredient to produce 
meaningful models is to have significant observational restrictions of high quality. 

Furthermore, a precise chemical evolution model for the MW is important 
because, by trying to reproduce the observed abundances \citep[particularly the new 
determinations achieved with 8-10 m-class telescopes; e.g.][]{garcia2006,martin2015,esteban2017}, 
we can put constraints on the many parameters involved in the formation and evolution of the  MW.

Most of the effort, of the chemical evolution models of the Galactic disk found in the literature, 
has been placed on fitting the observed radial distribution of $M_{\it gas}$, of $M_{\it stars}$, 
and of the SFR. The chemical gradients of HII regions have not been used as robust 
observational constraints because,  while the slope of the chemical gradient is not controversial,
the absolute O/H values of the HII region gradients found in the 
literature present a big dispersion; this becomes more pronounced when combined with gradients derived from other 
types of objects (such as B-stars, cepheids, PNe, etc.). Therefore, to constrain the chemical 
enrichment, some models fit the observed [O/Fe] - [Fe/H] trend shown by stars at the solar 
neighbourhood \citep[e.g.][]{renda2005,molla2015}, and others focus on normalizing their results 
to the solar abundances at the Sun's age \citep[e.g.][]{romano2010,minchev2013,prantzos2016}. 

It is known that the Sun might have migrated away from its birth galactocentric distance \citep[e.g.][]{wielen1996,portegies2009}
and detailed dynamical evolution models of the MW
provide a restriction for the birth place of the Sun given by $6.3 \lesssim R \lesssim 9.1$ kpc  \citep{martinez2017}

The O/H gradient is obtained from young objects assuming that they are representative
of their present R value. Therefore we use \ion{H}{2} regions, early type main sequence B stars,
and young Cepheids (all objects with ages smaller than 200 Myr) to determine the O/H
values. These objects are very young and we do not expect them to have experienced
important migrations since their birth. Unfortunately O/H values derived by different
techniques and authors show very large dispersions \citep[higher that 1.0 dex; see, for example, fig. 4 in][]{kubryc2015a}.

In the specific case of  the O/H gradients derived from \ion{H}{2} regions;
there are two different strategies to derive O/H values:  
a) those based on the direct method (DM) 
where the ${\rm O^{++}/H^+}$ values are derived from the forbidden lines of 
${\rm O^{++}}$, and the temperature of the ${\rm O^{++}}$ region from the ratio of the 
auroral (4363 \AA) to nebular (5007 \AA) emission lines \citep[e.g.][]{fernandez2017,esteban2018},  
and b) those based on a
temperature independent method (TIM) where the ${\rm O^{++}/H^+}$ values are 
derived from the recombination lines of ${\rm O^{++}}$ and ${\rm H^+}$ \citep[e.g.][]{esteban2005}.

The difference arises because the DM is very sensitive to the temperature and in the presence of temperature 
inhomogeneities only provides a lower limit to the O/H ratio 
\citep{peimbert1967,peimbert1969}.
A more accurate method to obtain the O/H ratio is based on the recombination lines (RLs) 
of O because their temperature and density dependences are
similar to those of the H RLs and nearly cancel out, 
making the O/H ratio practically temperature and density independent.

The problem with the TIM is that the intensities of the \ion{O}{2} RLs are difficult to obtain: 
a) they are about three to four orders of magnitude fainter than the [\ion{O}{3}] nebular 
lines and about one order of magnitude fainter than the [\ion{O}{3}] auroral lines, and 
b) to obtain accurate \ion{O}{2} intensities requires high spectral resolution; these 
difficulties imply that, to obtain TIM abundances, observations with large telescopes are 
required. Overall this method is seldom used and most determinations in the literature 
are based on O forbidden lines (FLs).

Alternatively, \citet{peimbert1967} and \citet{peimbert1969} presented a formalism to correct the chemical abundances determined using CELs in the presence of thermal inhomogeneites; this correction is possible for objects where the temperature dispersion can be estimated. This is called the $t^2$ formalism and the $t^2$ value is a measure of the temperature dispersion. One way to obtain a $t^2$ value is to compare 
FL temperatures with either \ion{H}{1} 
or \ion{He}{1} temperatures; unfortunately good \ion{H}{1} and \ion{He}{1} temperatures are 
available for only a few objects, usually only for those where \ion{O}{2} RL abundances are 
available.

Recent reviews on the determination of abundances based on the DM and on the TIM have 
been presented by \citet{perez2017} and \citet{peimbert2017} respectively.

One of the goals of this paper is to select carefully the data to have a smaller dispersion
on the O/H values and to obtain a meaningful O/H gradient. A special effort will be made
in the determination of the O/H abundance ratios in \ion{H}{2} regions because these objects are
the most recent and are not affected by migration; there are systematic differences in the 
O/H values that amount typically to about 0.25 dex between the two most popular methods; 
one issue that we are going to tackle is which of those two methods is better suited to model the MW.

The main objective of this paper is to present a new Galactic model of chemical evolution 
able to reproduce the best observational data; 
specifically a model capable of simultaneously fitting
the data provided by \ion{H}{2} regions, Cepheids, B stars, and the Sun;
such model will improve our understanding of the chemical evolution 
of the Galactic disk, specially of the last 5 Gyrs.

\section{O/H values determined from \ion{H}{2} regions }\label{direct}

We are looking for objects with high quality observations, preferably observed with 8-10m class telescopes;
we found 10 \ion{H}{2} regions for which oxygen RLs are measured, all of them observed with the Ultraviolet/Visible Echelle Spectrograph (UVES) at the 8.2m Very Large Telescope (VLT); since most of these stars are inside the solar circle, we have extended this sample with 9 \ion{H}{2} regions observed with long slit spectrographs with the 10.4m Gran Telescopio de Canarias (GTC), mostly outside the solar circle; we also included 2 \ion{H}{2} regions 
observed with the 4.2m William Herschel Telescope (WHT). The entire set is presented in Table 1.

Recently, \citet{esteban2018} published a sample with additional data in the internal part of the Galaxy.
However, the highest O$^{++}$ fraction of such sample is 0.27 dex while the sample by \citet{pena2012} 
only has two objects with O$^{++}$ below 0.27 dex and, since the determinations of Pe\~na-Guerrero et al. 
rely heavily on O$^{++}$ data,  the calibration of low ionization objects would be suspect; also this new 
sample is much fainter than the sample we are using (the typical errors in O/H for our sample are $\sim 0.04$ 
dex while the errors in the new sample are closer to 0.08 dex); consequently we have not used these data.

\subsection{Direct method}

In Table 1 we present the O/H gaseous abundances derived from forbidden lines under the 
assumption that there are no temperature inhomogeneities, ${\rm (O/H)_{FL}}$. We also 
present the correction to these abundances due to the fraction of O atoms trapped in dust 
grains in \ion{H}{2} regions, ${\rm (O/H)_{FL} + DUST}$. The fraction of O trapped into dust 
grains depends slightly on metallicity, and for these chemical abundances is in the 0.10 to 
0.11 dex range \citep{peimbert2010}.

In Figure 1a we present two straight lines to fit the data. For the 4 to 10 kpc range the slope 
amounts to $-0.0170$ dex/kpc, while for the 9.5 to 17.5 kpc range the slope amounts to 
$-0.0629$ dex/kpc. In Figure 1b we present a parabolic fit to the O/H values derived from 
the DM. The parabolic fit for the log O/H versus $R$ data is better than the linear one. The 
equation for the parabolic fit is:
\begin{equation}
12 + {\rm log {O/H}} =  8.571 + 0.0325 \left({{R} \over {\rm kpc}}\right) - 0.00353  \left({{R} \over {\rm kpc}}\right)^2.
\end{equation}

\subsection{Temperature independent method}

In Table 2 we present the O/H gaseous abundances derived from RLs, ${\rm (O/H)_{RL}}$. 
We also present the correction to these abundances due to the fraction of O atoms trapped 
in dust grains  ${\rm (O/H)_{RL} + DUST}$ \citep{peimbert2010}. They are the best observed \ion{H}{2} regions 
and comprise a set of 10 objects.

For objects where no RLs are available, no true TIM abundances can be obtained, however it is 
possible to correct for an average abundance discrepancy factor, ADF, for objects where only FL abundances are available 
\citep[e.g.][and references therein]{pena2012,peimbert2017}. 

In Table 3 we present a set of 11 \ion{H}{2} regions where in column 3 we include again 
the ${\rm (O/H)_{FL}}$ values for these objects. Also in column 4 of this Table we include 
the ${\rm (O/H)_{TOTAL}}$ values where we have adopted the calibration by 
\citet{pena2012} to correct for the presence of dust and temperature inhomogeneities in 
this set of objects. The correction by Pe\~na-Guerrero et al. was determined by fitting the 
dependence of the ADF to both O/H and O$^{++}$/O; in that 
work it was found that the correction has a weak dependence with metallicity and a 
negligible dependance with degree of ionization. For this set of objects corrections are 
in the  0.19 to 0.22 dex range.

In Figure 2 we include the set of 10 objects where the ${\rm (O/H)_{TOTAL}}$ values 
were obtained from the RLs, and the set of 11 objects where the ${\rm (O/H)_{TOTAL}}$ 
were obtained from the ${\rm (O/H)_{FL}}$ values and the calibration by \citet{pena2012}. 

In Figure 2a we present the O/H values for the galactic \ion{H}{2} regions based on the 
TIM.  Also in Figure 2a we present two pairs of lines to fit the data  for the 4 to 10 kpc 
range the slope  amounts to $-0.0263$ dex/kpc  and  for the 9.5 to 17.5 kpc range the 
slope amounts to  $-0.0638$ dex/kpc. In Figure 2b we present a parabolic fit to the O/H 
values derived from the TIM. The parabolic fit for the log O/H versus $R$ data is better 
than the linear one. The equation for the parabolic fit is:
\begin{equation}
\label{equation:TIM}
12 + {\rm log (O/H)} =  8.888 + 0.0142 \left({{R} \over {\rm kpc}}\right) - 0.00289  \left({{R} \over {\rm kpc}}\right) ^ 2.
\end{equation}
Based on this equation, it is possible to compare the observed slopes by other authors 
and compare them with our observations (see Table 4).

From Figures 1 and 2 and equations (1) and (2) it can be seen that \ion{H}{2} regions show a 
flattening of the O/H gradient that increases from the outer parts to the inner parts of the 
Galaxy, the 18 to 6 kpc range.

\section{O/H values determined from other objects}\label{Other values}

\subsection{Comparison with the  early B type stars in the Orion Nebula}

From 13 B0 V to B2 V stars of the Orion OB 1 association \citet{nieva2011} obtain that 
$12 + {\rm log(O/H)} = 8.77 \pm 0.03$; later on, from B-stars located at distances from 
the Sun smaller than 350 pc, \citet{nieva2012} find $12 + {\rm log(O/H)}= 8.76 \pm 0.05$. 
Both results are far from the 8.59 dex value obtained by the DM, but  in excellent 
agreement with the 8.80 obtained from equation (2) and the $8.76 \pm 0.04$ value obtained 
from the TIM for the Orion nebula by \citet{esteban2004}. 

\subsection{Early B type stars}

In Figure 3 we present a set of 18 O/H values in the 6-18 kpc range obtained from 51 
early type B-stars of Galactic, open cluster associations by \citet{rolleston2000}. The 
distances to these associations are well known and their stars are younger than 100 Myr, 
therefore these abundances should be representative of the present ISM values. 
We also present 4 B-stars, studied by the same group, with galactocentric distances in the 2-5 kpc 
range \citep{smartt2001}.

\subsection{Cepheids}

The Cepheids used in this paper are Type I or classical, which are younger than 
200 Myr, and consequently can be compared, both chemically and dynamically, with 
\ion{H}{2} regions and main sequence early B type stars.

In  Figure 3  we present also a set of 397 disk Cepheids compiled by Martin
 (private communication, 2017) based on data by his Andrievsky's group
\citep{luck2013,korotin2014,martin2015,andrievsky2016}.
From this figure  we conclude that 
the O/H values derived with the TIM are in reasonable agreement with the O/H 
values of the Cepheids and B-stars, on the other hand the O/H values derived from 
the DM are typically from 0.2 to 0.3 dex smaller than those derived with early 
type B-stars and disk Cepheids. 

\subsection{Combined Fit}

 Since the three sets of O/H data: \ion{H}{2} regions, early B-stars, and Cepheids are in reasonable agreement, we also present a fit which combines the three sets to have an observational constraint on the O/H values for different galactocentric distances. The most direct way of doing so is to simultaneously fit all the observed data points 
regardless of their origin; this would be a good approach if all the data points had the same origin, or had the same systematic errors. However, we are dealing with three different sets of data that probably present different systematics; also there are enough data points in the samples so the errors should be dominated by the systematic errors within each method and not due to the statistical errors within the sample. Therefore we consider that each data set should be treated independently; and then the fits should be averaged giving each set an equal weight.

In Figure 4 we present the O/H data 
provided by Cepheids, B-stars, and \ion{H}{2} regions using the TIM. We also present 
the parabolic fit to the three sets of data giving to each set one third of the weight. 
The parabolic weighted fit is given by:
\begin{equation}
12 + {\rm log (O/H)} =  8.952 + 0.0033 \left({{R} \over {\rm kpc}}\right) - 0.00245  \left({{R} \over {\rm kpc}}\right) ^ 2.
\end{equation}
This fit is very similar to that given by equation (\ref{equation:TIM}), and since this paper 
is centered in \ion{H}{2} regions we will use equation (\ref{equation:TIM}) to fit the 
chemical evolution models.

\section{Chemical evolution models for the disk of the Galaxy}\label{models}

We present three chemical evolution models for the Galactic disk. All these models were built to 
reproduce three present-time observational constraints along the Galactic disk: the 
distributions of the total baryonic mass, $Mtot(R)$, the gas mass, $Mgas(R)$,
and the O/H values for \ion{H}{2} regions between $5<R(kpc)<17$.

Based on observations, $Mtot(R) = 50e^{-(R-8)/Rdisk}$ \citep{fenner2003},  where the disk scalelength, 
$Rdisk$, ranges between 2.3 and 5 kpc for different spectral bands \citep{yin2009}; for our models, 
$Rdisk$ is a free parameter determined to reproduce the slope of the O/H gradient for $R > 12$ kpc. The observed 
$Mgas(R)$ values are obtained by adding the atomic and molecular data from fig. 7 by 
\citet{kennicutt2012}. These $Mgas(R)$ values include the hydrogen and helium components.

The age of the models is 13 Gyr and the Sun formed at $t= 8.4$ Gyr in the solar vicinity,
which is located in a ring of 8-kpc-medium radius and with an 0.5-kpc width. The MW 
disk started to form at $t =1$ Gyr in an inside-out scenario of primordial infall with time-scales
$\tau_{disk}= (R/kpc - 2)$ Gyr  \citep{chiappini1997}. The halo formed during the first Gyr of the evolution with a 
constant timescale $\tau_{halo} = 0.5$ Gyr (see Figure 5).

The star formation rate is a spatial and temporal function of the form
$SFR(R, t) = \nu(R,t) \times Mtot^{0.4}(R, t) \times Mgas^{1.4}(R, t)$, where 
$Mtot=Mgas +Mstar$  \citep{matteucci1999}. The values of $\nu(R,t)$ that we choose are those that reproduce the 
general behavior of $Mgas(R)$ and the flattening of the O/H radial gradient for $R \le 5$ 
kpc. The initial mass function is that by \citet{kroupa1993} in the $0.08 - Mup$ \msun  \ range,
where $Mup$ is a free parameter, that is obtained by matching the absolute value of the observed O/H 
gradient.

The metal-dependent yields for $m < 8 $ \msun  \ were taken from 
\citet{marigo1996,marigo1998} and \citet{portinari1998} and for $m > 8$ \msun \ from 
\citet{hirschi2007}, \citet{meynet2002}, and \citet[][high mass loss rate model]{maeder1992}.
Since the Fe yields are not computed by the Geneva group, we adopted the 
\citet{woosley1995} values, following the \citet{carigi2008} prescription. For binary stars 
($3 < mbin$(\msun)$<16$), we adopted the yields by \citet{nomoto1997} for type Ia 
supernovae by using the formulation by \citet{greggio1983}. We considered $A_{bin} = 0.066$, 
as the fraction of binary stars that become SNIa progenitors.

The three models focus on $ R> 3$ kpc, therefore the Galactic bar and bulge are not considered,
and no galactic winds are assumed.

Two of these models (CP11 and TIM) were built to reproduce the observed O/H values 
based on the temperature independent method determinations (Figures 6 and 7)
and the third model (DM) was built to reproduce the O/H values of the \ion{H}{2} regions derived from the direct method
(Figures 8 and 9).

\subsection{Astrophysical choices for the ingredients of the models}

As most representative works on chemical evolution models for spiral galaxies,
we adopt an inside-out formation scenario for the Galactic disk. 
\citet{molla2016} computed 
different infall rates for 16 theoretical galaxies with virial masses between $5 \times 10^{10}$ 
and $10 \times 10^{13}$ \msun; they concluded that the initial infall rate is greater, yet 
decreases more rapidly, in the inner disk; thus agreeing with the inside-out escenario and 
with cosmological simulations. In a recent work, \citet{nuza2018} studied the infall $R$ and 
$t$ dependencies in 4 simulated MW-like galaxies; they found that the inner disks assemble 
in shorter times than  the outer disks, as the inside-out scenario predicts. Also, the metallicity 
distribution function observed by APOGEE at different $R$s \citep{hayden2015} suggest an 
inside-out formation.

In our models, we consider the well known Kennicutt-Schmidt law, 
$ {\it SFR}(R,t) \propto Mgas^{1.4}(R, t)$, obtained empirically for spiral galaxies \citep{kennicutt1998}.
The proportionality coefficient we use is given by $(Mgas +Mstar)^{0.4}$, and can be approximated by 
$Mstar^{0.4}$ (since $Mgas << Mstar$ for most of the evolution). Recently \citet{shi2018} 
presented a revision of the Schmidt law based on spatially resolved SFR, Mgas, and Mstar from dwarf, 
merging, and spiral galaxies. They found  the star formation efficiency in galaxies depends on the 
stellar mass surface density, as $ Mstar^{0.5}$, very similar to ours. 

\citet{romano2005} tested several IMFs and concluded that the Kroupa et al. (1993) IMF 
reproduces many important constraints of the solar vicinity, and hence, is the one used in our models. \citet{romano2010} 
showed a comparative study among different stellar yields, and concluded that the high wind yields 
improve the model predictions. Moreover, they mentioned other independent and successful chemical 
evolution models that considered yield sets for massive stars similar to those assumed in our CEMs.

More recently \citet{berg2016} studied the C/O-O/H trend in dwarf galaxies together with that 
shown by stars of the solar vicinity.  They compared these trends with 
the C/O-O/H relation predicted by independent CEMs that considered different combinations of stellar 
yields and IMFs. In that work, Berg et al. concluded that ``The Carigi \& Peimbert model [our CP11 model]
is the most successful at reproducing the general trend seen for the collective log(C/O) data over the 
observed range in oxygen abundance".

\subsection{CP11 Model}

The CP11 model corresponds to the high wind yield model by \citet{carigi2011}.
They considered the yields computed from high stellar wind for massive stars of high metallicity. 
Based on the O/H data from the TIM, obtained before 
2011 and that only included objects in the 6-11 kpc range, Carigi and Peimbert concluded 
that the best model was the one that assumed a moderate mass loss rate for massive stars 
of solar metallicity, given by the average of the high and low wind models. However, the 
high wind model reproduces better the new O/H data derived by the TIM presented in this 
paper (See Figure 5). The high wind yield model by CP11 considered $Rdisk=3.5$ kpc, 
$Mup=80 $ \msun,  and $\nu=0.019$ as a spatial and time constant. 
\citet{carigipeim2008,carigi2011} presented very good fits to the high wind model
with the C/H and C/O gradients as well as with the C/O-O/H, C/Fe-Fe/H, O/Fe-Fe/H, and time-Fe/H 
evolutionary trends at the solar vicinity.
In Figures 6, 7, and 10 
it can be noted the good agreements of this model with the current observational data.
In Figure 11 (left panels)  we show the evolution of the O/H values and the SFR.

\subsection{TIM model}

This model is identical to the CP11 model, except on the star formation history for $R<6$ kpc.
To reproduce the O/H flattening (12+log(O/H) $\sim 8.83$, see Figure 6) we assume a unit step function, such that  diminishes abruptly the {\it SFR(R)} at specific times  (quenching times) as a function of  $R$. 
Then we adopt the current star formation rate observed at these 
radii \citep{kennicutt2012} for the rest of the evolution. 
For $R=$ 3, 4, and 5 kpc, the observed 
SFR is equal to 0.45, 0.55, and 0.65 \msun pc$^{-2 }$Gyr$^{-1}$, respectively, see Figure 10.
We ran several models, testing different quenching times, where we chose the highest times that match 12+log(O/H) $ \sim 8.83$.
In Figure 11 (central panels), the decrease of the SFR and its effect on O/H can be noted,
specifically, for $R=$ 3, 4, and 5 kpc the quenching times are 4.08, 6.10, and 8.38 Gyr, respectively.

In Figure 6 we present the model built to reproduce the O/H values of \ion{H}{2} regions 
derived from the TIM and corrected by dust depletion. For $R<6$ kpc the TIM model 
presents a flat gradient emulating a quasi flat behavior shown by \ion{H}{2} regions. 
This flattening is a consequence of the assumptions of an inside-out reduction of the star 
formation rate. The abundances predicted by the TIM model 
only differ from those by CP11 in the 3-6 kpc range. For comparison, we also plot the 
abundances from the CP11 model.

In Figure \ref{Model-CELs}a we compare the O/H radial distribution in the interstellar 
medium at present time, obtained by the TIM and CP11 models, with B-stars of Galactic 
open cluster associations by \citet{rolleston2000} at different $R$, with the O/H average 
of B-stars in the solar vicinity by \citet{nieva2012}, as well as with individual B-stars near 
the centre of the Galaxy ($2.5 \le R \le 4.7$ kpc) from \citet{smartt2001}. Since these stars 
are of spectral type B0 V to B2 V, they have been recently formed; and, consequently, 
their chemical abundances should be representative of  the current O/H values in the 
interstellar medium. We find that in general both models are in good agreement with the 
observations, however the TIM model produces a slightly better fit for $R<6$ kpc,
since this model is in agreement, at one sigma, with all the values for $R \le 4.7$ kpc.

In Figure \ref{Model-CELs}b we compare the current O/H radial distribution from both models, 
with the Cepheid data. As B-stars, Cepheids are young stars (younger than 200 Myr), and we 
expect their abundances to be representative of the current ISM. We find that the values 
predicted by the TIM and CP11 models are similar to those of these stars. 
However, for the inner disk, the CP11 
model matches better the few observations for $R < 6$ kpc.
More data are needed to  discriminate between both models.

In Figure \ref{Model-CELs}c we present the predicted O/H radial distribution at 8.4 Gyr, 
the evolutionary-time when the Sun was formed. We consider that the age for the Sun is 4.6 Gyr 
and the solar initial O/H value is the protosolar value, $12 + {\rm log (O/H)} = 8.73 \pm 0.05$, 
by \citet{asplund2009}. The galactocentric distance at which the Sun was born is not known, 
but recent dynamical models for the solar orbit predict a value of $7.7 \pm 1.4$ kpc at one 
sigma for the solar origin \citep{martinez2017}. The spatial, temporal and O/H agreements 
with the model are excellent. Therefore, the model does not need to invoke a radial migration 
of the Sun to explain its chemical properties. 

To estimate the birth radius of the Sun,  we compare the protosolar abundance, 
$12 + {\rm log (O/H)} = 8.73 \pm 0.05$ \citep{asplund2009}, with the chemical abundances 
produced by our models 4.6 Gyr ago. Specifically, O/H abundances of 8.68 and 8.78 are 
achieved at $R=8.7$ and $R=6.3$ kpc in both the TIM and CP11 models, see Figure 7c.
Moreover, the dynamic models by \citet{martinez2017} indicate that the Sun may have been 
born between 6.3 and 9.1 kpc. Therefore, we estimate that the Sun originated between 6.3 
and 8.7 kpc and that the Sun may have migrated between 1.7 kpc outwards and 0.7 kpc 
inwards, to its present location.

Our determination of the birth radius of the Sun is in agreement, at one sigma, with those 
obtained from chemodynamical models by \citet{minchev2013}(4.4 to 7.7 kpc) and by 
\cite{kubryc2015a} (6.8 kpc) but slightly higher than the value mentioned by \citet{nieva2012} 
(5 to 6 kpc).

\subsection{DM Model}

Since the ratios between O/H values determined from TIM and DM are nearly constant 
(i.e. there is no clear radial trend in the observed ADFs), we apply the same idea and procedure
of quenching for both models.
Because the O/H flattening value is lower from the DM, and the SFR is higher at the inner galactocentric radii from the inside-out scenario, we impose a SFR quenching at earlier times for the DM model.
Specifically, to reproduce the O/H flattening (12+log(O/H) $\sim 8.7$) for R = 3, 4, and 5 kpc, the quenching times are 3.30, 3.50 and 5.40 Gyr, respectively.
See Figures. 8, 9, 10, and 11 (right-bottom panel).

Since the O/H values from the DM are lower  by $\sim 0.25$ dex than those from the TIM model,
the DM model needs a smaller amount of stars to produce O, consequently an 
$Mup=40$ \msun \ is required. Similarly to obtain a better match for the O/H values with 
$R >12$ kpc a value for $Rdisk = 4.5$ kpc is required.

Similarly to the previous section, in Figures 8 and 9, we present the results of the model 
built to reproduce the O/H values of the \ion{H}{2} regions derived from the DM.

The flat behavior shown by \ion{H}{2} regions in the 5 to 8 kpc range (see Figure 8), 
can be theoretically reproduced by an earlier inside-out reduction of the star formation 
rate (see Figure 11). In Figure 9 (panels a and b) we compare the current O/H vs $R$ 
relation from the model with the O/H values from B-stars and Cepheids, respectively. 
It can be noted that the agreement between the observed O/H values and the model 
is poor, being about 0.25 dex lower than observed. In Figure 9c, the predicted initial 
solar value, 12 + log  O/H, is in the 8.40 to 8.56 range, about 0.25 dex smaller 
than the initial solar value that amounts to 8.73 \citet{asplund2009}. No migration 
(outwards or inwards) can explain the O/H solar value, without reducing the solar age. 
In an extreme case, had the Sun been born at $R \sim 6$ kpc,  the O/H value of the 
ISM at $R=6$ kpc reaches $\sim 8.73$ dex  at $t \sim$12 Gyr (see Figure 11), this 
would imply a solar age of  $\sim 1$ Gyr, in contradiction with the accepted value 
of 4.6 Gyr.

\section{Discussion}\label{Disc}

A fundamental problem to make successful CEMs is to choose which observational data to fit to the models.
In this work we are focusing on the long standing problem of whether  DM
or TIM abundances should be used as observational restrictions. 
\citet{robles2013} find for M33 that, on average, \ion{H}{2} region DM abundances are $\sim 0.34$ dex smaller than those of B-supergiants; while \ion{H}{2} region abundances corrected by $t^2$ (equivalent to the TIM) are consistent with the stellar and RL abundances. Alternatively, \citet{bresolin2009} find, for NGC 300, that CEL abundances agree better with abundances derived from AB supergiants. \citet{toribio2016} observe both galaxies and find, again, a better agreement for RLs in M33 and a better agreement  for CELs in NGC 300. \citet{bresolin2016} suggest that DM abundances agree better with the AB supergiant abundances at sub-solar metallicities, while at super-solar metallicities RLs agree better with the supergiant data.  Our result agrees with Robles-Valdez et al.; we find that DM abundances fall short of stellar abundances in the MW for \ion{H}{2} regions between 5 and 17 kpc (equivalent to $8.3\lesssim12+$log(O/H)$\lesssim 8.88$). 

All these observations should be used to restrict the several inputs and free parameters of the CEMs.
The chemical history of the CP11 and TIM models have been successfully tested in the 
solar vicinity \citep[see][]{carigipeim2008,carigi2011} because in our neighborhood the number of 
observational constraints is similar to the number of free parameters of the CEMs, (i.e., infall, SFR, 
IMF, and stellar yieds; see Section \ref{models}). 


As we mention in the Section \ref{models}: the upper mass limit, $Mup$, of the initial mass 
function is a free parameter of the chemical evolution models, that is obtained mainly 
from the absolute \ion{H}{2} region O/H values. Since the O/H values in the 
Galactic disk determined from the TIM are $\sim$ 0.25 dex higher than those  
obtained from the DM, and O is produced mainly by massive stars, the CP11 and 
TIM models require an $Mup=80$ \msun, while the DM model requires an 
$Mup=40$ \msun.

Similar $Mup$ values were found by \citet{hernandez2011} for NGC~6822, an irregular 
galaxy of the Local Group. Their chemical evolution model built to reproduce O/H values 
determined from recombination lines (or TIM, in the present paper), required an 
$Mup=80$ \msun, while their model computed by matching the O/H values from 
collisionally excited lines (or DM, in the present paper), required an $Mup=40$ \msun.

The absolute value of current O/H$(R)$ could be reproduced by CEMs that use stellar yields and IMFs different 
to those assumed in this paper; nevertheless, those CEMs could not reproduce the chemical history of the solar 
vicinity, that is, the abundance ratios shown by stars of diverse ages in our neighborhood, including the Sun. 
Moreover, with those sets, the star formation history may change, and consequently the fitting to the metal 
distribution functions would change.
In any case, even if one assumes other sets of yields and IMFs, other $Mup$ values would be required, but
always $Mup({\it TIM}) > Mup({\it DM})$, because the O/H values determined from the TIM are higher than 
those from the DM.

\citet{molla2015} tested stellar yields and IMFs using 144 CEMs, all built to reproduce the solar 
abundances. They found 4 best models that reproduce: i) for the Galactic disk, the radial 
distribution of the gas mass, stellar mass,  SFR, C/H, N/H, and O/H; and ii) for the solar vicinity, 
the SFR and [Fe/H] evolutions, the [$\alpha$/Fe] vs [Fe/H] trend and the metal distribution 
function, both for [Fe/H]$ < 0$. Unfortunately, they cannot achieve neither the [$\alpha$/Fe] nor 
the [Fe/H] values observed in the most metal rich stars ([Fe/H]$ > 0$) of the solar vicinity, because 
the O/H gradient used as observational constraint presents a mean dispersion of $\sim$ 0.6 dex 
at the solar galactocentric radius. It is worth to note that  all their best models consider the IMF 
by Kroupa (2002) \citep[identical for $m > 0.08$ \msun to the][IMF used in our models]{kroupa1993}.
 
\citet{mishurov2018} have produced a Galactic disk model to fit the Cepheids O/H 
values obtained by \citet{martin2015}. Their model is in agreement with the 
\ion{H}{2} region O/H values derived with the TIM abundances
in the 7-14 kpc range. For $R < 7$ kpc their predicted O/H values become 
higher than the observed ones, as much as $\sim 0.3 dex$.

In this work, we propose a simple formula of inside-out quenching of the SFR, as one of the processes that 
produces an O/H flattening; there are other, more complex, formulas able to produce a similar drop in 
the inner O/H gradient. We need ages and chemical abundances of stars located in the inner galactocentric 
radii to confirm the flattening and the validity of this quenching formula.  Moreover, cosmological simulations and 
dynamical models could give some hints to other flattening formulas and other possible alternatives.
However, our simple formula can reproduce the color gradients present in local galaxies 
\citep[e.g.][]{lian2017} and in spiral galaxies located in the transition region between  star forming galaxies and  quiescent ones \citep{belfiore2017}.

In a recent work, \citet[][see also references therein]{esteban2018} mentioned some possible reasons behind 
the star formation quenching: a gas depletion in the inner disk induced by the Galactic bar, or an increase 
of the gas turbulence within the bar-corotation radius ($3.4 - 7$ kpc). However, they compared the flattening 
radius of the O/H gradient in the MW ($R \sim 6$ kpc) with the inner-drop radius in the O/H gradient of other 
galaxies; this drop occurs at about one half the effective radius (that corresponds to $R \sim 2$ kpc for the MW);
although 6 kpc is within the estimated value for the corotation radius of the MW bar, it is much larger than what 
is observed in other galaxies \citep{sanchez2018}.

In particular, between 3 and 6 kpc, we increased the amount of accreted gaseous mass, 
and did not incorporate this extra fresh gas to the SFR. 
For these models to reproduce the O/H, the predicted current gas mass in the inner disk, 
is higher than observed (by $\sim 0.2$ and 0.3 dex for values from TIM and DM methods, respectively).

To test for alternatives to explain the O/H flattening we explored dilution models 
without changing the star formation history. Extreme dilution models can produce any flattening we require,
 since an accretion of primordial gas can increase the H abundance significantly.
In particular, between 3 and 6 kpc, we increased the amount of accreted gaseous mass, 
 and did not incorporate this extra fresh gas to the SFR. For these models to reproduce 
 the O/H, the predicted current gas mass in the inner disk, is higher than observed, by about a 
 factor of 2, and the required infall at present time is more than 3 orders of magnitude larger than what is observed.

Based on the current set of O/H data from \ion{H}{2} regions, B-stars, and young Cepheids we 
cannot conclude if the gradient for $R < 6$ kpc is steep or flat. For this reason, we present 
the CP11 and TIM models, the first one was built to reproduce a steep inner gradient and
the second model, to match a flat gradient. The TIM model explains the O/H flattening as 
a consequence of an abrupt decline of the star formation rate from 3 to 5 kpc, and from 
$\sim 4.1$ to 8.4 Gyr, as a consequence of an inside-out star formation quenching.

Some recent chemical and chemodynamical models for the Galactic disk do not seem 
to produce a flat inner O/H gradient, by considering radial migration of gas and/or stars 
induced mainly by the Galactic bar. \citet{cavichia2014} obtained a very steep O/H 
gradient in the 4 - 8 kpc range, contrary to the observations. \citet{minchev2014} predicted 
a steep O/H gradient for $R < 10$ kpc, even steeper than the CP11 gradient for $R < 5$ kpc.
\citet{kubryc2015b} achieved a flat gradient, but only for the 3 to 4 kpc range. 

\section{Summary and Conclusions}\label{Conc}

The temperature independent method (the one based on the recombination lines of O and H),
produces O/H values that are systematically higher than those derived from the direct method
(the one  based on the temperature derived from the ratio of the auroral to nebular
lines of O). The abundance discrepancy between these 2 methods can be explained by the 
existence of temperature inhomogeneities, that undermine the basic assumptions of the DM 
determinations producing lower O/H values than the TIM ones.

The correction of the O/H values due to the temperature inhomogeneities is about 2 times 
higher than the correction due to the fraction of O embedded in dust grains.

A parabolic fit in the 5 to 18 kpc range is better than a linear fit for the log O/H versus $R$ 
distribution. Therefore we expect a flatter O/H gradient in the inner parts than in the outer 
parts of the disk for \ion{H}{2} regions and for recently formed stars, those with ages smaller 
than about 0.5 Gyr.

We present two galactic chemical evolution models, one that fits the O/H values in \ion{H}{2}
regions based on the DM, and another that fits the O/H values based on the TIM. 
We also discuss the model by \citet{carigi2011} that is in excellent
agreement with the model that fits the O/H values based on the TIM
for the 6 to 18 kpc range.

The current O/H abundances derived from the TIM chemical evolution model
are in excellent agreement with the observations of: a) Cepheids, b) early B type stars, 
and c) the Sun (after modeling the chemical evolution of the Galaxy since the 
Sun was formed). The ability to reproduce the chemistry present when the Sun was 
formed, as well as the current chemical gradient show the capacity of the TIM model to 
reproduce the last 5~Gyrs of galactic chemical evolution.

On the other hand, the current O/H abundances  derived from the DM 
chemical evolution model are typically smaller by about 0.25 dex than those observed 
in: a) Cepheids, b) early B type stars, and  c) the Sun (after modeling the chemical 
evolution of the Galaxy since the Sun was formed), regardless of the stellar migration 
assumed for the Sun.

As expected, the TIM model presented in this paper is
in excellent agreement with the abundances derived in the 6 to 18 kpc range. 
However, more abundance determinations are needed in the 3 to 6 kpc range to discriminate between 
the CP11 and TIM models.

If the flattening of the O/H ratio range is corroborated, it would imply an
inside-out star formation rate quenching in the 3-5 kpc range starting $\sim$9 Gyrs ago.


We are grateful to R. Pierre Martin, C\'esar Esteban and Sebasti\'an S\'anchez for pertinent 
letters and many fruitful discussions. 
We are also grateful to the referee for many relevant suggestions that improved this paper.
We would like to acknowledge support from CONACyT, 
grant 247132; L.C. also received partial support from PAPIIT (DGAPA-UNAM), grants no. 
IG100115, IA101215, and IA101517) as well as partial support from MINECO (Spain), 
grant no. AYA2015-65205-P; A.P acknowledge support from PAPIIT (DGAPA-UNAM), 
grant no. IN109716.

\clearpage

\begin{deluxetable}{lr@{$\pm$}lr@{$\pm$}lr@{$\pm$}ll} 
\tablecaption{Forbiddeen O/H abundances for $t^2 = 0.00$ and including dust
\label{foha}}
\tablewidth{0pt}
\tablehead{
\colhead{Name}  & 
\multicolumn{2}{c}{$R$ (kpc)} &
\multicolumn{2}{c}{${\rm O/H _{FL}}$} &
\multicolumn{2}{c}{${\rm O/H _{FL}+DUST}$\tablenotemark{a}} &
\colhead{Source\tablenotemark{b}} }
\startdata
M 20          &  5.1 &   0.3 &  8.51 & 0.04 & 8.62 & 0.05 & (1, 2) \\
M 16          &  5.9 &   0.2 &  8.54 & 0.04 & 8.65 & 0.05 & (1, 2) \\  
M 17          &  6.1 &   0.2 &  8.54 & 0.04 & 8.65 & 0.05 & (1, 3) \\
M  8           &  6.3 &   0.8 &  8.45 & 0.04 & 8.56 & 0.05 & (1, 3) \\
NGC 3576 &  7.5 &   0.3 &  8.55 & 0.04 & 8.66 & 0.05 & (1, 4) \\
IC 5146     & 8.10 & 0.02 & 8.56 & 0.04 & 8.67 & 0.05 & (1 ,5) \\
M 42          & 8.34 & 0.02&  8.50 & 0.04 & 8.61 & 0.05 & (1, 6) \\
NGC 3603 &  8.6 &   0.4 &  8.44 & 0.03 & 8.55 & 0.04 & (1, 2) \\
Sh 2-100   &   9.4 &  0.3 &  8.52 & 0.03 & 8.63 & 0.04 & (1) \\
Sh 2-132    & 10.0 & 0.7 &  8.35 & 0.14 & 8.46 & 0.14 & (1, 7) \\
NGC 7635 & 10.2 &  0.7 &  8.40 & 0.08 & 8.51 & 0.08 & (1, 8) \\
Sh 2-156   & 10.6 &  0.6 &  8.32 & 0.10 & 8.43 & 0.10 & (1, 7) \\
Sh 2-311   & 11.1 &   0.4 &  8.39 & 0.01 & 8.50 & 0.03 & (1, 9) \\
Sh 2-298   & 11.9 &  0.7  &  8.41 & 0.02 & 8.52 & 0.03 & (1) \\
NGC 2579 & 12.4 &  0.5 &  8.26 & 0.03 & 8.36 & 0.04 & (1, 10) \\
Sh 2-128   & 12.5 &  0.4 &  8.19 & 0.03 & 8.29 & 0.04 & (1) \\   
Sh 2-288   & 14.1 &  0.4 &  8.31 & 0.08 & 8.42 & 0.08 & (1) \\
Sh 2-127   & 14.2 &  1.0 &  8.25 & 0.04 & 8.35 & 0.05 & (1) \\
Sh 2-212   & 14.6 &  1.4 &  8.15 & 0.12 & 8.25 & 0.12 & (1) \\
Sh 2-83     & 15.3 &  0.1 &  8.14 & 0.05 & 8.24 & 0.06 & (1) \\
Sh 2-209   & 17.0 &  0.7 &  8.09 & 0.10 & 8.19 & 0.10 & (1)
\enddata
\\
\tablenotetext{a}{The dust contribution is based  on \citet{peimbert2010}, \citet{pena2012}), and \citet{espiritu2017}.}
\tablenotetext{b}{(1) \citet{esteban2017}; sources for ${\rm  (O/H)_{FL}}$ values: (1) \citet{esteban2017}, (2) \citet{garcia2006}, 
(3) \citet{garcia2007}, (4)\citet{garcia2004}, (5) \citet{garcia2014}, (6) \citet{esteban2004}, (7) \citet{fernandez2017}, (8) \citet{esteban2016}, 
(9) \citet{garcia2005}, (10) \citet{esteban2013}. }
\end{deluxetable} 

\clearpage

\begin{deluxetable}{lr@{$\pm$}lr@{$\pm$}lr@{$\pm$}ll} 
\tablecaption{Recombination O/H abundances corrected for dust
\label{roha}}
\tablewidth{0pt}
\tablehead{
\colhead{Name}  & 
\multicolumn{2}{c}{$R$ (kpc)} &
\multicolumn{2}{c}{${\rm O/H _{RL}}$} &
\multicolumn{2}{c}{${\rm O/H _{RL}+DUST}$\tablenotemark{a}} &
\colhead{Source\tablenotemark{b}}}
\startdata
M 20            &      5.1  & 0.3 &              8.71 & 0.07 &             8.82  & 0.08   &              (1)  \\
M 16              &    5.9  & 0.2   &            8.81  & 0.07  &            8.92  & 0.08    &             (1) \\
M 17                &   6.1  & 0.2    &           8.76  & 0.04   &          8.87  & 0.05      &           (2) \\
M  8              &      6.3  & 0.8      &         8.71 & 0.04      &        8.82  & 0.05        &          (2) \\
NGC 3576      &    7.5  & 0.3        &       8.74  & 0.06       &      8.85  & 0.07          &        (3) \\
M 42             &     8.34 & 0.02        &   8.65  & 0.03   &           8.76  & 0.04             &     (4) \\
NGC 3603    &      8.6  & 0.4   &           8.72  & 0.05    &          8.83  & 0.06              &    (1) \\
Sh 2-100         &   9.4  & 0.3      &        8.58  & 0.05       &        8.69 & 0.06       &           (5) \\
Sh 2-311     &      11.4  & 0.4        &     8.57  & 0.05          &     8.68  & 0.06         &         (6) \\
NGC 2579     &    12.4  & 0.5         &    8.56  & 0.03           &   8.67  & 0.05           &       (7)
 \enddata
\\
\tablenotetext{a}{The dust contribution is based  on \citet{peimbert2010}, \citet{pena2012}), and \citet{espiritu2017}.}
\tablenotetext{b}{(1) \citet{garcia2006}, (2) \citet{garcia2007}, (3)\citet{garcia2004}, (4) \citet{esteban2004}, 
(5) \citet{esteban2017}, (6) \citet{garcia2005}, (7) \citet{esteban2013}. }
\end{deluxetable} 

\clearpage

\begin{deluxetable}{lr@{$\pm$}lr@{$\pm$}lr@{$\pm$}l} 
\tablecaption{Forbidden O/H abundances corrected for dust and temperature inhomogeneities
\label{fohat2}}
\tablewidth{0pt}
\tablehead{
\colhead{Name}  & 
\multicolumn{2}{c}{$R$ (kpc)} &
\multicolumn{2}{c}{${\rm O/H _{FL}}, t^2 = 0.00$} &
\multicolumn{2}{c}{${\rm O/H _{TOTAL}}$} }
\startdata
IC 5146  &          8.10 & 0.02&         8.56 & 0.04&           8.89 & 0.06 \\
Sh 2-132  &       10.0 & 0.7     &      8.35 & 0.14   &        8.76 & 0.15 \\
NGC 7635  &     10.2 & 0.7       &    8.40 & 0.08     &      8.72 & 0.09 \\
Sh 2-156       &   10.6 & 0.6         &  8.32 & 0.10       &    8.63 & 0.11 \\
Sh 2-298         & 11.9 & 0.7            &8.41 & 0.02         & 8.73 & 0.05 \\
Sh 2-128&          12.5 & 0.4&            8.19 & 0.03&          8.49 & 0.06 \\
Sh 2-288  &        14.1 & 0.4  &          8.31 & 0.08  &        8.62 & 0.09 \\
Sh 2-127    &      14.2 & 1.0    &        8.25 & 0.04    &      8.56 & 0.06 \\
Sh 2-212      &    14.6 & 1.4      &      8.15 & 0.12      &    8.45 & 0.13 \\
Sh 2-83          &  15.3 & 0.1        &    8.14 & 0.05        &  8.44 & 0.07 \\
Sh 2-209          &17.0 & 0.7          &  8.09 & 0.10          &8.38 & 0.12 \\
  \enddata
\\
The $R$ and ${\rm O/H _{FL}}$ values for $ t^2=0.00$ come from Table 1. 
The ${\rm O/H _{TOTAL}}$ values  include the $t^2$ effect and 
the dust correction based on the calibration by \citet{pena2012}.
Therefore the ${\rm O/H _{TOTAL}}$  values are equivalent to the 
${\rm O/H _{RL}+DUST}$ values.
\end{deluxetable} 

\clearpage

\begin{deluxetable}{lcccl} 
\tablecaption{Comparison of O/H gradients for different radial intervals
\label{grad}}
\tablewidth{0pt}
\tablehead{
\colhead{$R$ Interval}  & 
\colhead{${\Delta log(\rm O/H)}/\Delta R$} &
\colhead{Equation (2)}  & 
\colhead{Object} &
\colhead{Source} \\
\colhead{(kpc)}  & 
\colhead{other papers} &
\colhead{this paper} 
}
\startdata
  6 --11   & $-0.044 \pm 0.010$ & $-0.0349$ & \ion{H}{2} Regions & (1) \\
 11--17   & $-0.046 \pm 0.017$ & $-0.0654$ & \ion{H}{2} Regions & (2) \\                          
 11--18   & $-0.053 \pm 0.009$ & $-0.0678$ & \ion{H}{2} Regions & (3) \\                          
  5 --18   & $-0.060                 $ & $-0.0531$ & \ion{H}{2} Regions & (4) \\
  6 --18   & $-0.067                 $ & $-0.0556$ & B Stars                  & (5) \\
  5 --18   & $-0.058                 $ & $-0.0531$ & Cepheids               & (6) \\
  5 --17   & $-0.056                 $ & $-0.0507$ & Cepheids               & (7) \\
 \enddata
\\
(1) \citet{esteban2005}, (2) \citet{esteban2017}, (3) \citet{fernandez2017}, (4) \citet{rudolph2006}, (5) \citet{rolleston2000}, (6) \citet{korotin2014}, (7) \citet{luck2011}.
\end{deluxetable} 

\clearpage


\begin{figure}
\begin{center}
\includegraphics[angle=0,scale=0.7]{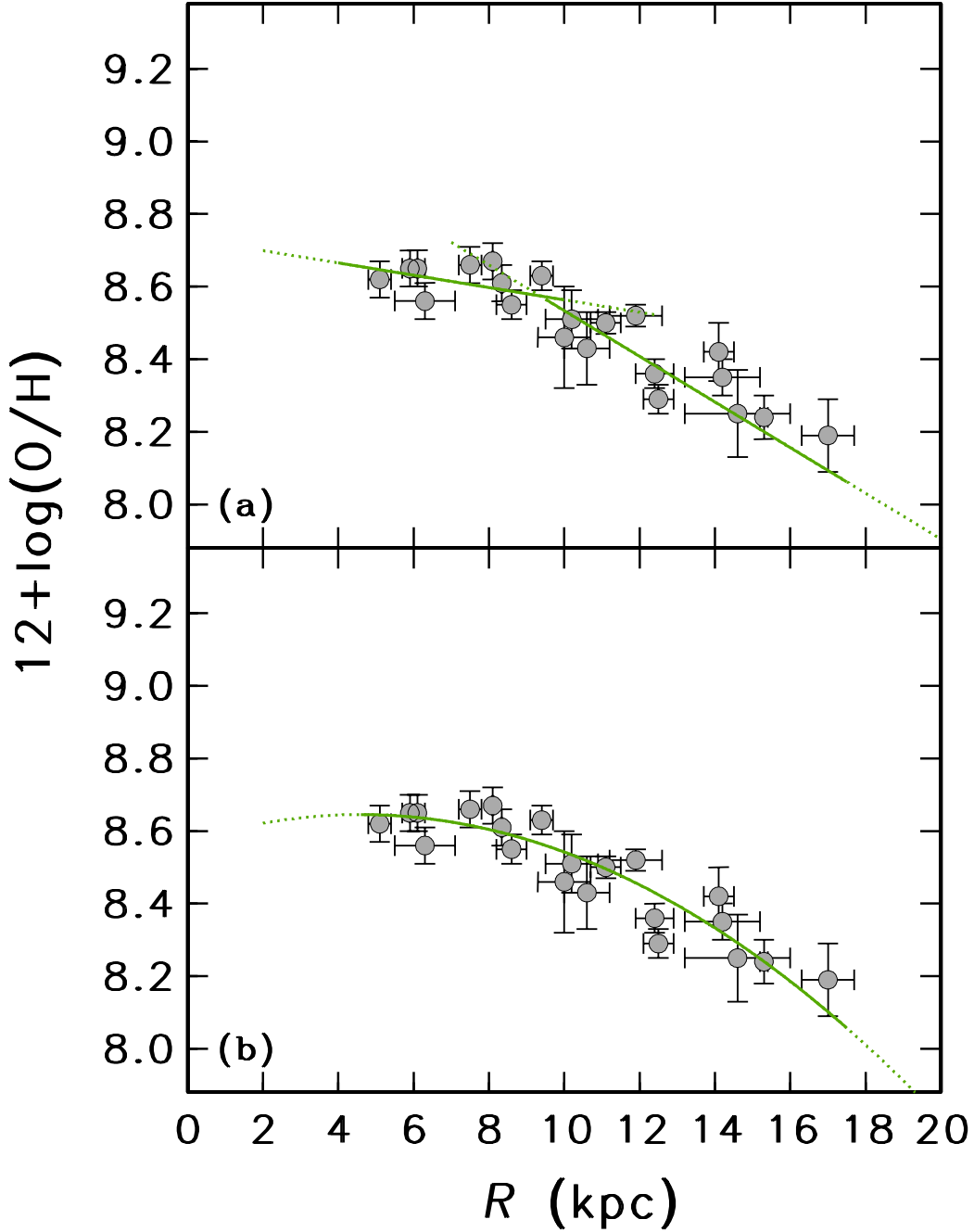}
\end{center}
\caption[f1.eps]{
\ion{H}{2} region O/H values obtained with the DM plus dust corrections versus the radial 
distance to the galactic center, $R$ (see Table 1).  The O/H values have been corrected 
by the fraction of O embedded in dust grains. In panel a the straight lines are the fits to 
the 4 - 10 kpc  and the 9.5 - 17.5 kpc ranges and the slopes are: 0.0170 and 0.0629 dex/kpc, 
respectively. In panel b, a parabolic fit to the O/H values is shown the solid line corresponds 
to the analytic fit where data is available, the dotted lines correspond to extrapolations.
\label{CELs-2lines-parabola}}
\end{figure}

\clearpage

\begin{figure}
\begin{center}
\includegraphics[angle=0,scale=0.7]{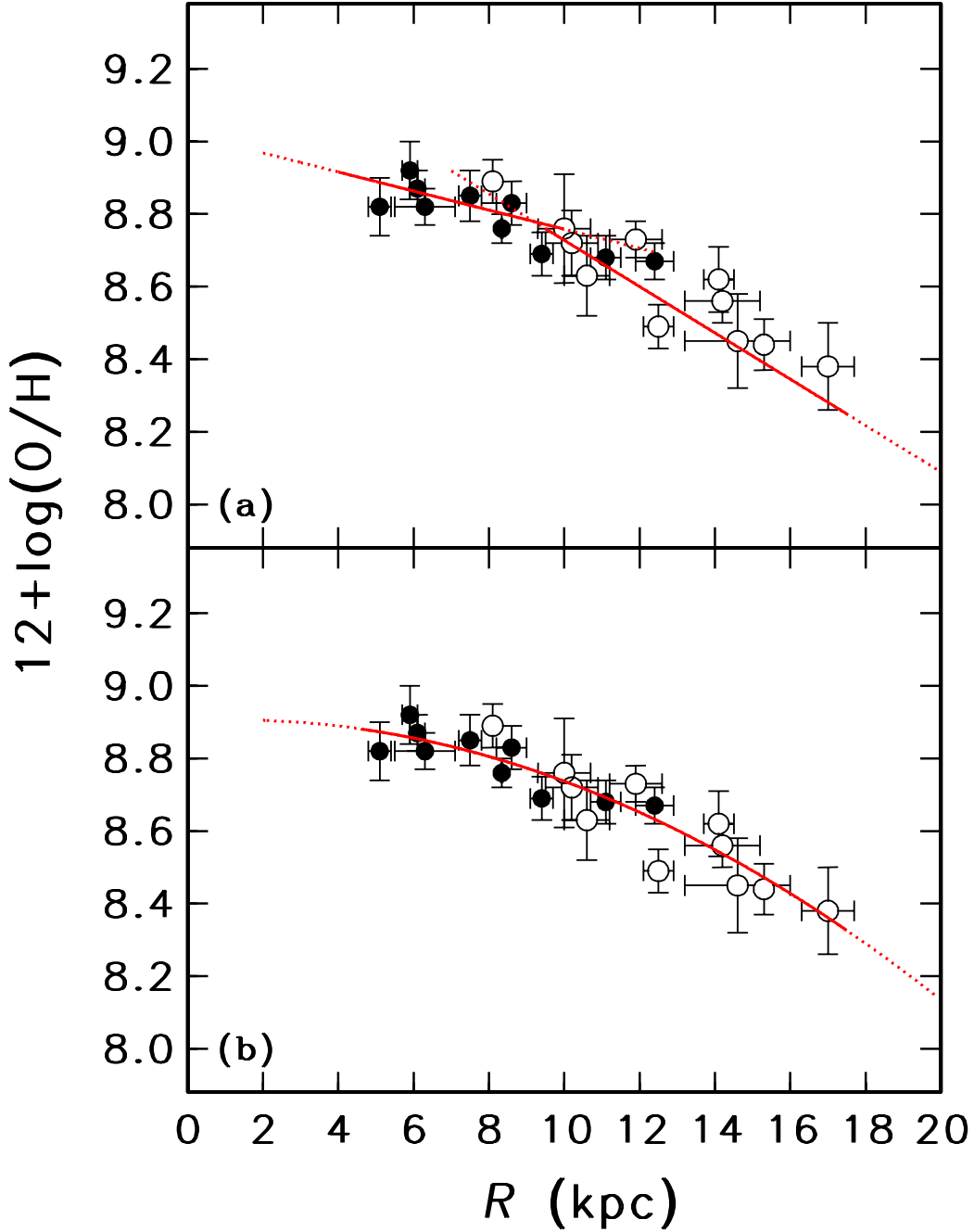}
\end{center}
\caption[f2.eps]{
\ion{H}{2} regions O/H values obtained with the TIM plus dust corrections versus the radial 
distance to the galactic center, $R$ (see Table 3). The O/H values include the fraction of 
O embedded in dust grains. The straight lines are the fits to the 4 - 10 kpc  and the 
9.5 - 17.5 kpc ranges and the slopes are: 0.0263 and 0.0638, respectively. Parabolic fit to 
the O/H values obtained with the TIM. The filled circles denote direct observations of the 
recombination lines, the empty circles use the calibration by \citet{pena2012}. The solid 
line corresponds to the analytic fit where data is available, the dotted lines correspond to 
extrapolations.
\label{RLs-2lines-parabola}}
\end{figure}

\clearpage

\begin{figure}
\begin{center}
\includegraphics[angle=270,scale=0.625]{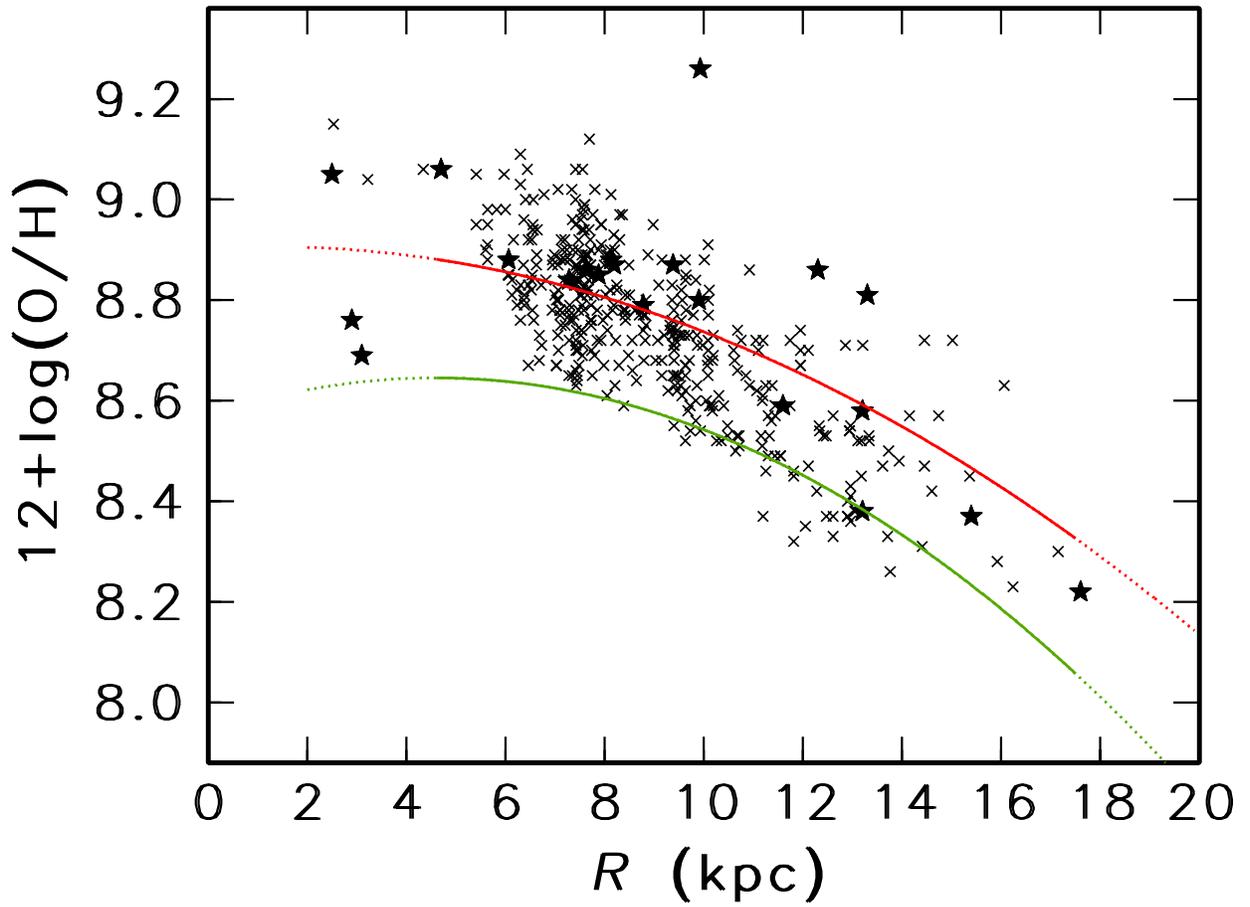}
\end{center}
\caption[f3.eps]{
Parabolic fits of O/H for the DM (green) and the TIM (red) plus dust corrections together with the 
B-star data  \citep[stars;][]{rolleston2000, smartt2001} and the Cepheid data (crosses; 
Martin 2017, private communication, see Section 3.3).
\label{parabola-CELs-vs-RLs}}
\end{figure}

\clearpage

\begin{figure}
\begin{center}
\includegraphics[angle=270,scale=0.625]{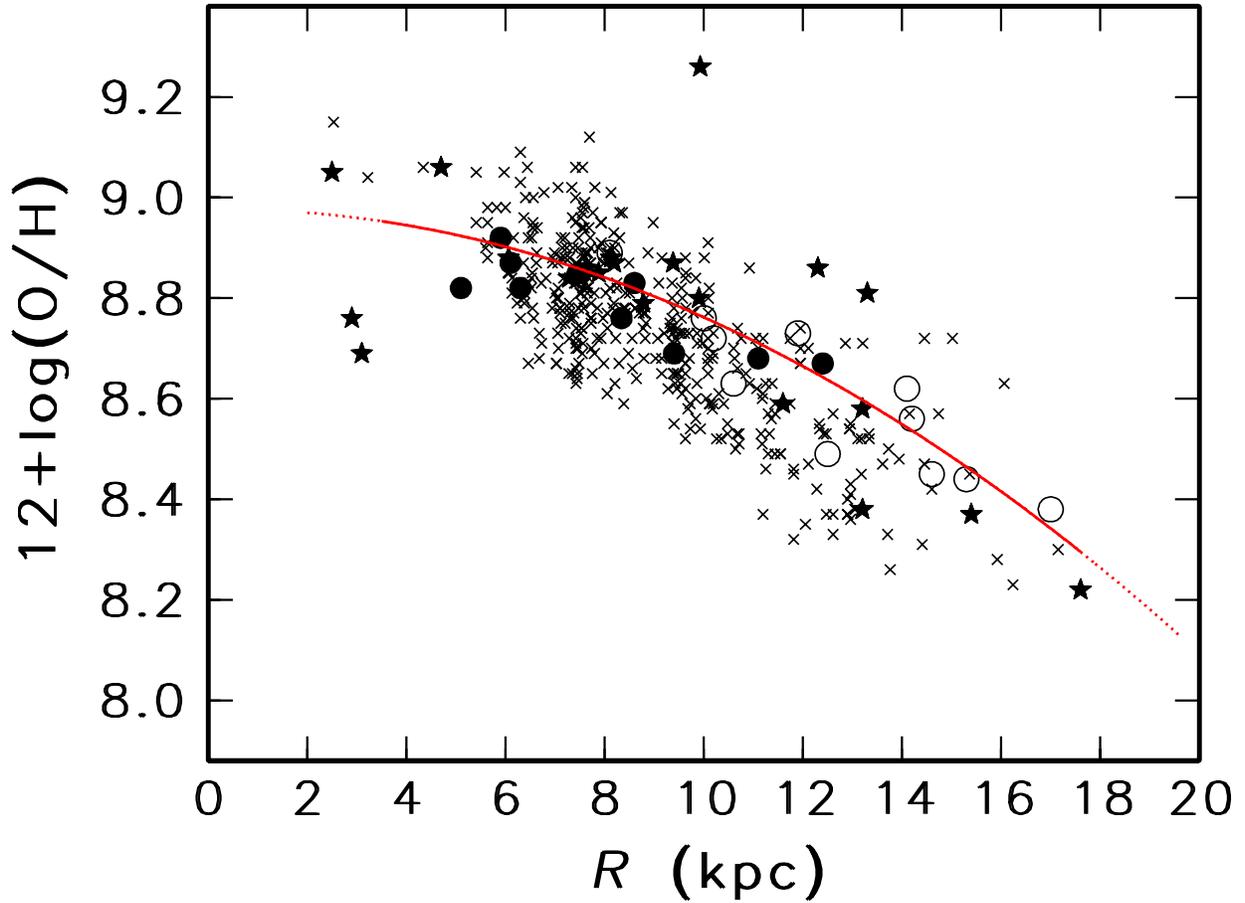}
\end{center}
\caption[f4.eps]{
Parabolic fit of O/H averaging fits to the \ion{H}{2} regions (circles; solid from RLs, empty 
from CELs corrected for $t^2$) plus dust corrections, to the B-star data 
\citep[stars;][]{rolleston2000, smartt2001}, and to the Cepheid data (crosses; Martin 2017, 
private communication, see Section 3.3).
\label{parabola-RLs-stars-cepheids}}
\end{figure}

\clearpage
\begin{figure}
\begin{center}
\includegraphics[angle=0,scale=0.825]{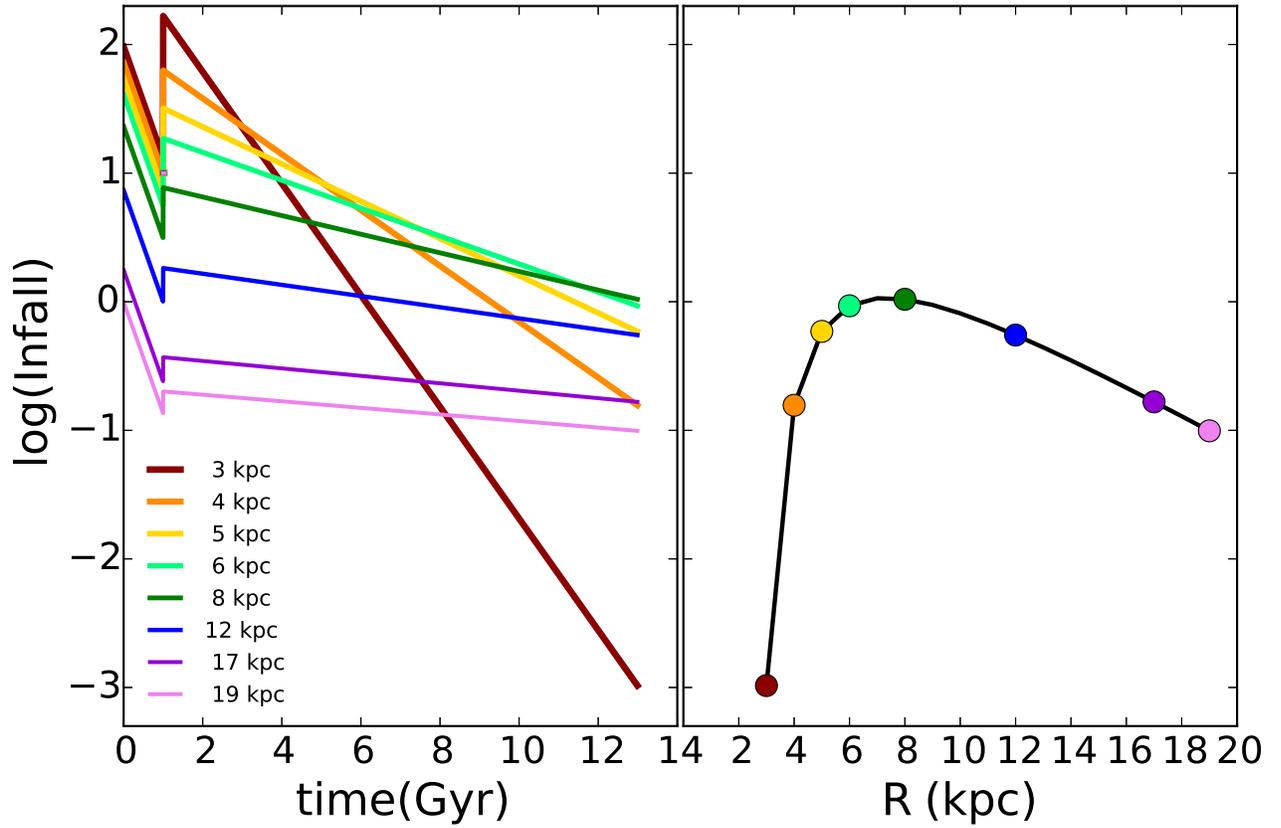}
\end{center}
\caption[f10.eps]{
Evolution for different radii and present-day radial distribution of the infall, for the three 
models shown in this paper. Circles represent the infall at present day (13 Gyr). Colour 
circles as colour lines. Infall in \msun pc$^{-2 }$Gyr$^{-1}$ units.
\label{Apendix3}}
\end{figure}

\clearpage
\begin{figure}
\begin{center}
\includegraphics[angle=0,scale=0.875]{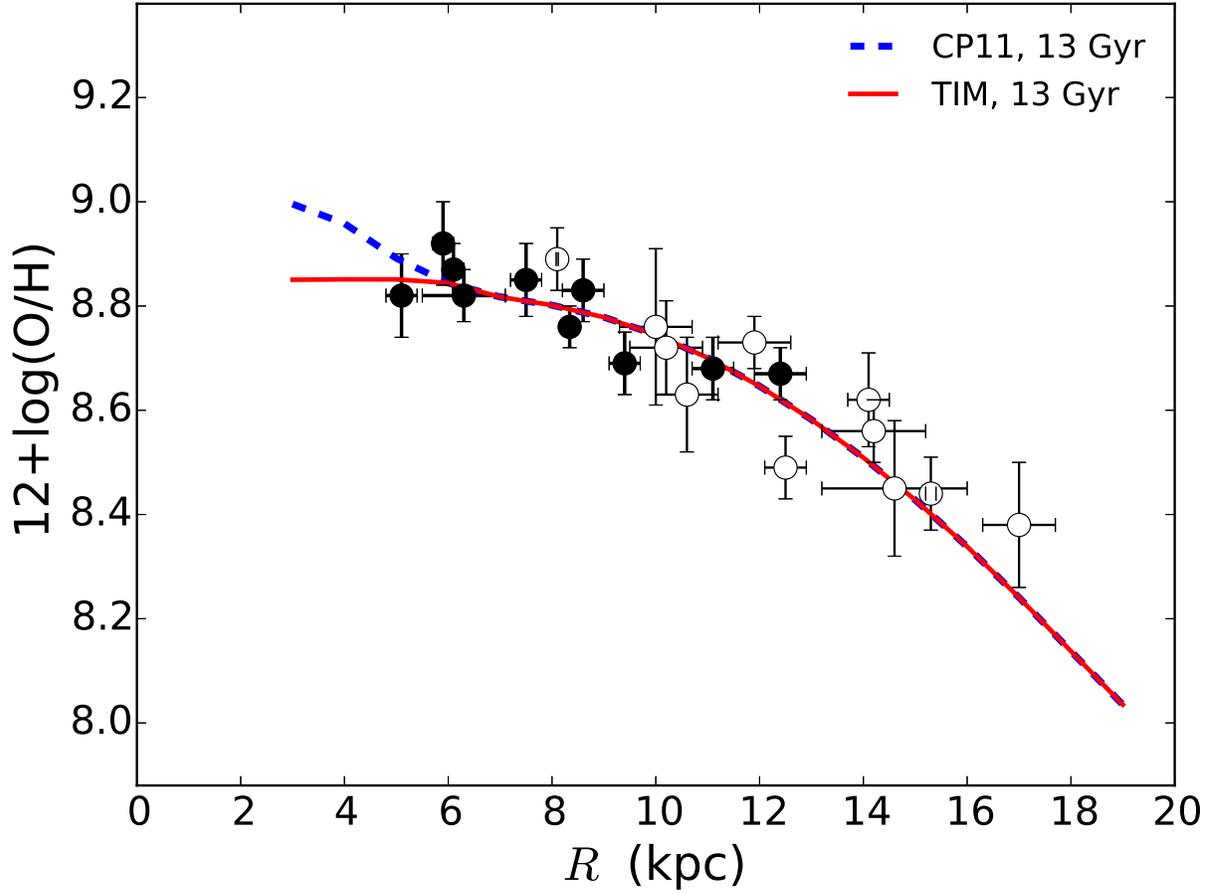}
\end{center}
\caption[f5.eps]{
Current radial distribution of O/H obtained from the Galactic chemical evolution model 
by CP11 and  from the model based on O/H values derived from the TIM plus dust 
corrections.  Data as in Figure 2. CP11 model was made based only on the data obtained 
before 2012, but it provides an excellent fit to the data by \citet{esteban2017} if one 
adopts the TIM.
\label{model-carigi2011}}
\end{figure}

\clearpage
\begin{figure}
\begin{center}
\includegraphics[angle=0,scale=1.000]{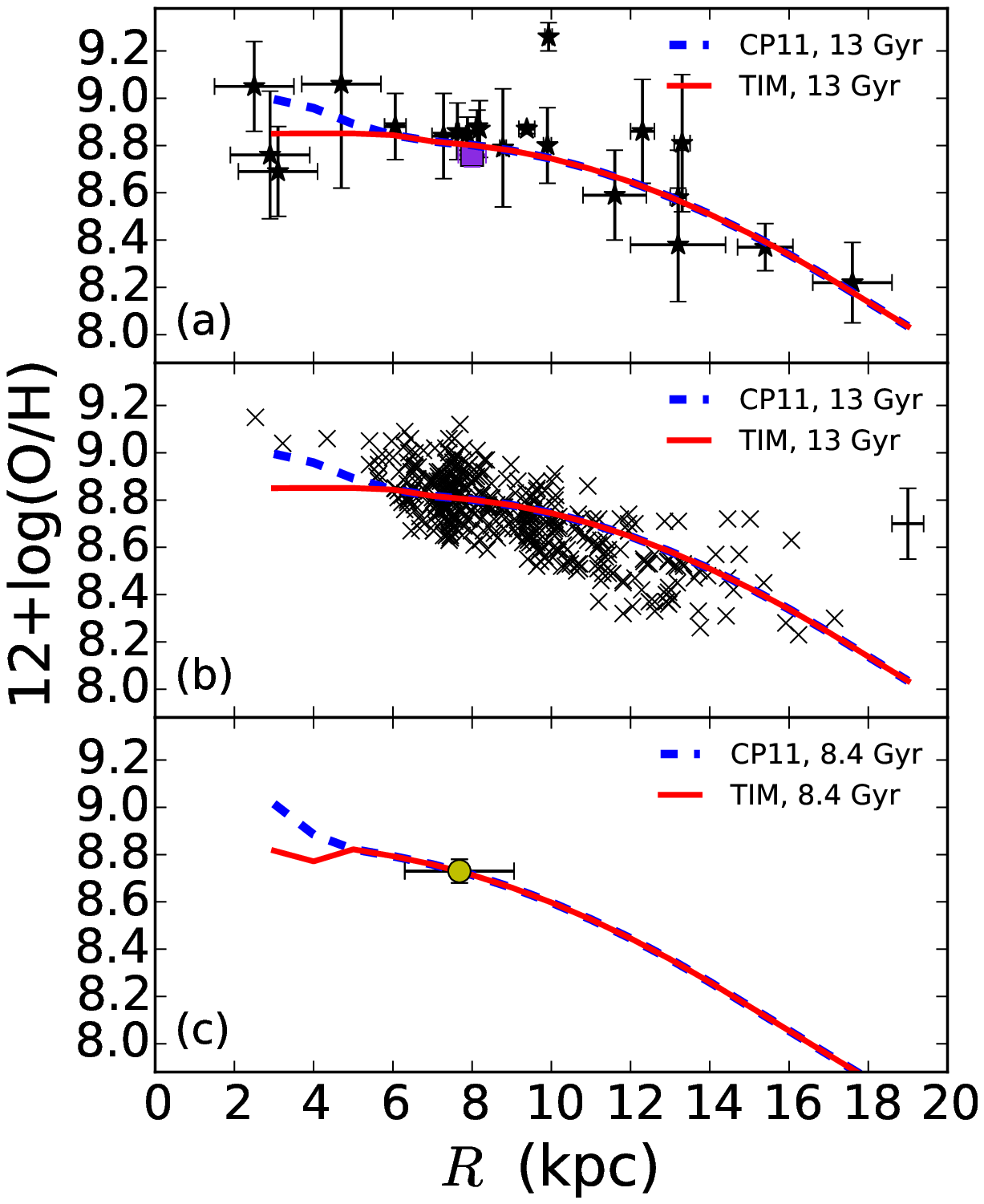}
\end{center}
\caption[f7.eps]{
Radial distribution of O/H at present (13 Gyr) and at the Sun formation time (8.4 Gyr) obtained 
by CP11 and TIM models. Panel (a): B-stars by \citet{rolleston2000} and \citet{smartt2001}, 
the violet square represents the average value and its standard deviation from B-stars by 
\citet{nieva2012}. Panel (b):  Cepheids as in Figure 3, the error bars presented near 19 kpc 
represent the uncertainty for the typical Cepheid. Panel (c): initial solar O/H value by 
\citet{asplund2009} filled circle, the horizontal error bar in the O/H solar abundance represents 
the average migration of the Sun computed by \citet{martinez2017}.
\label{Model-CELs}}
\end{figure}

\clearpage
\begin{figure}
\begin{center}
\includegraphics[angle=0,scale=0.875]{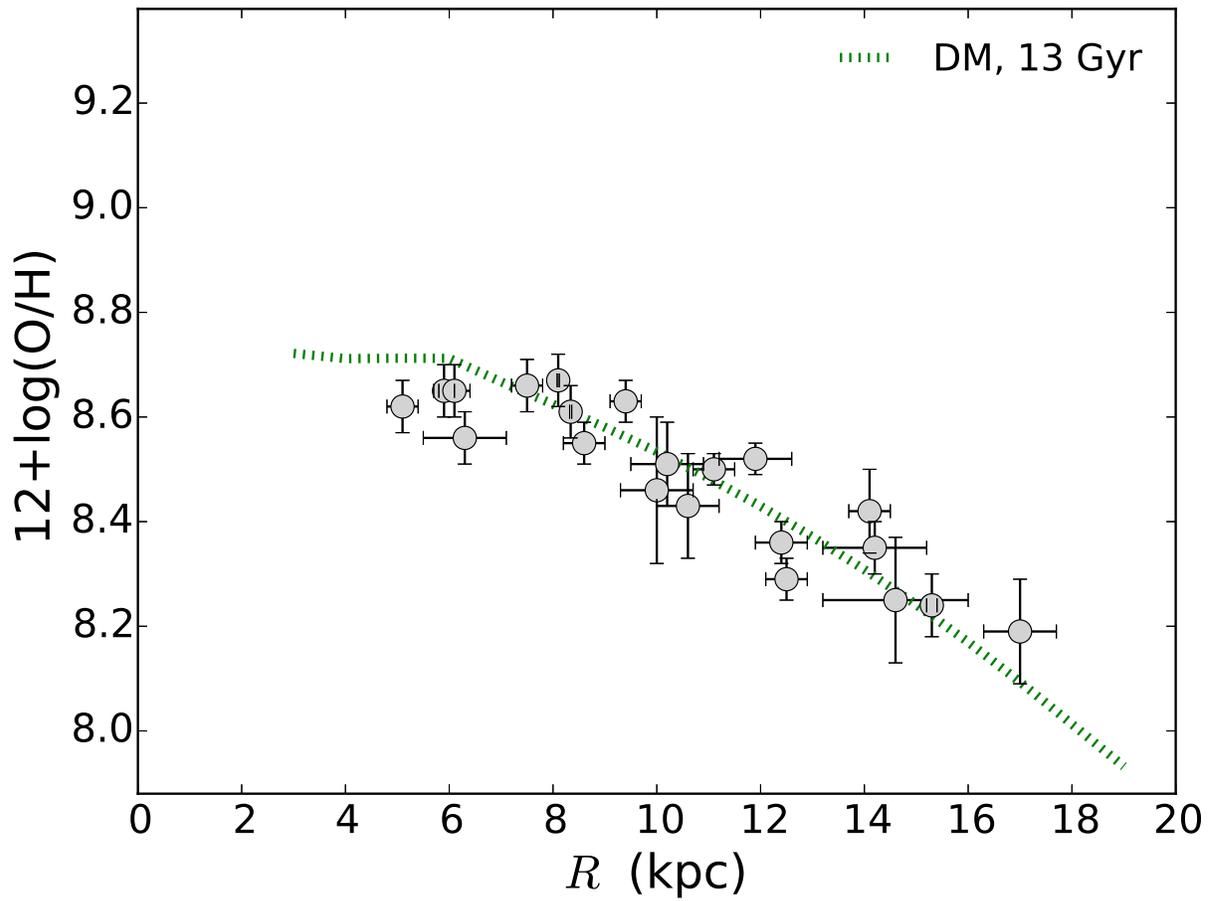}
\end{center}
\caption[f6.eps]{
Current radial distribution of O/H obtained from the Galactic chemical evolution model based 
on the O/H values derived from the DM. Data as Figure 1.
\label{DM-HII}}
\end{figure}

\clearpage

\begin{figure}
\begin{center}
\includegraphics[angle=0,scale=1.000]{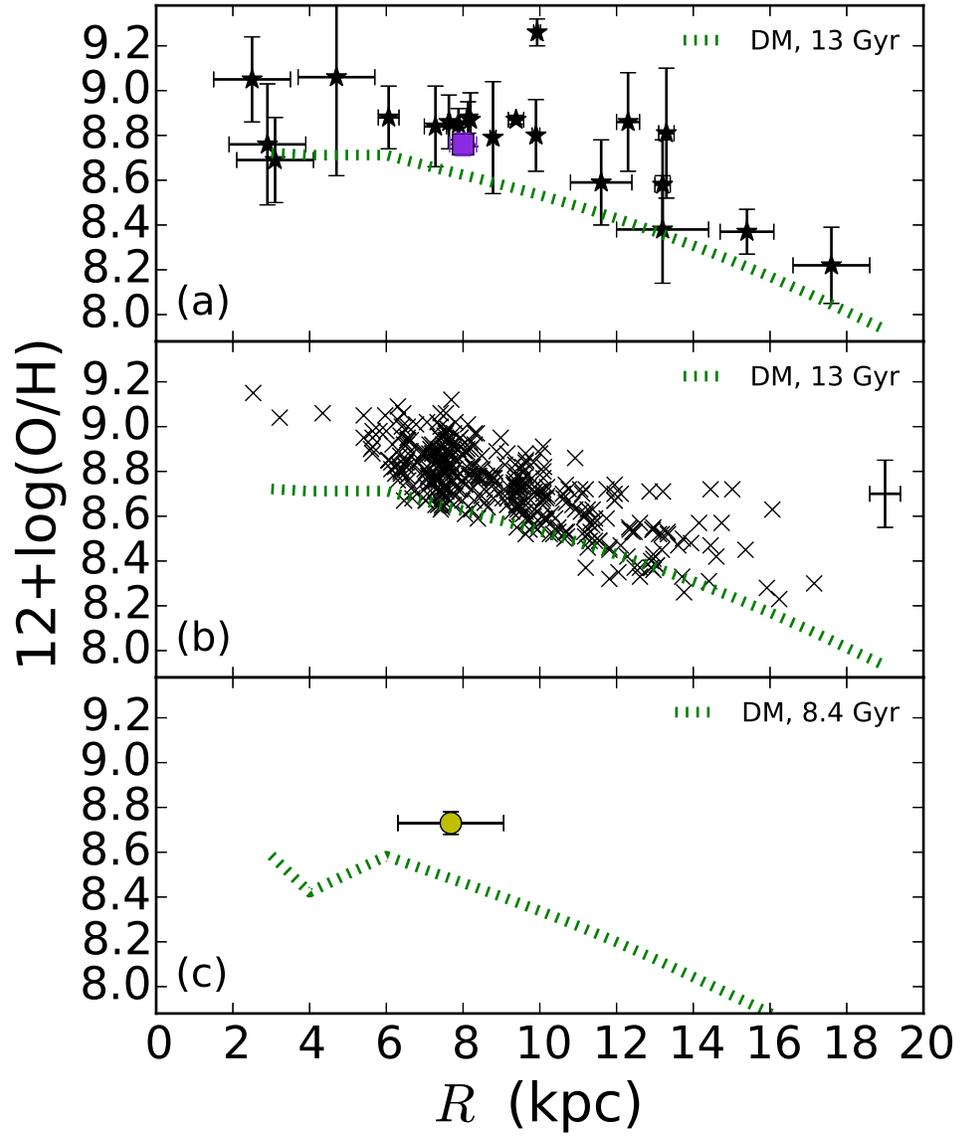}
\end{center}
\caption[f8.eps]{
Same data as in Figure 7, but now  compared with the model based on the DM.
\label{DM-Sun}}
\end{figure}

\clearpage

\begin{figure}
\begin{center}
\includegraphics[angle=0,scale=1.000]{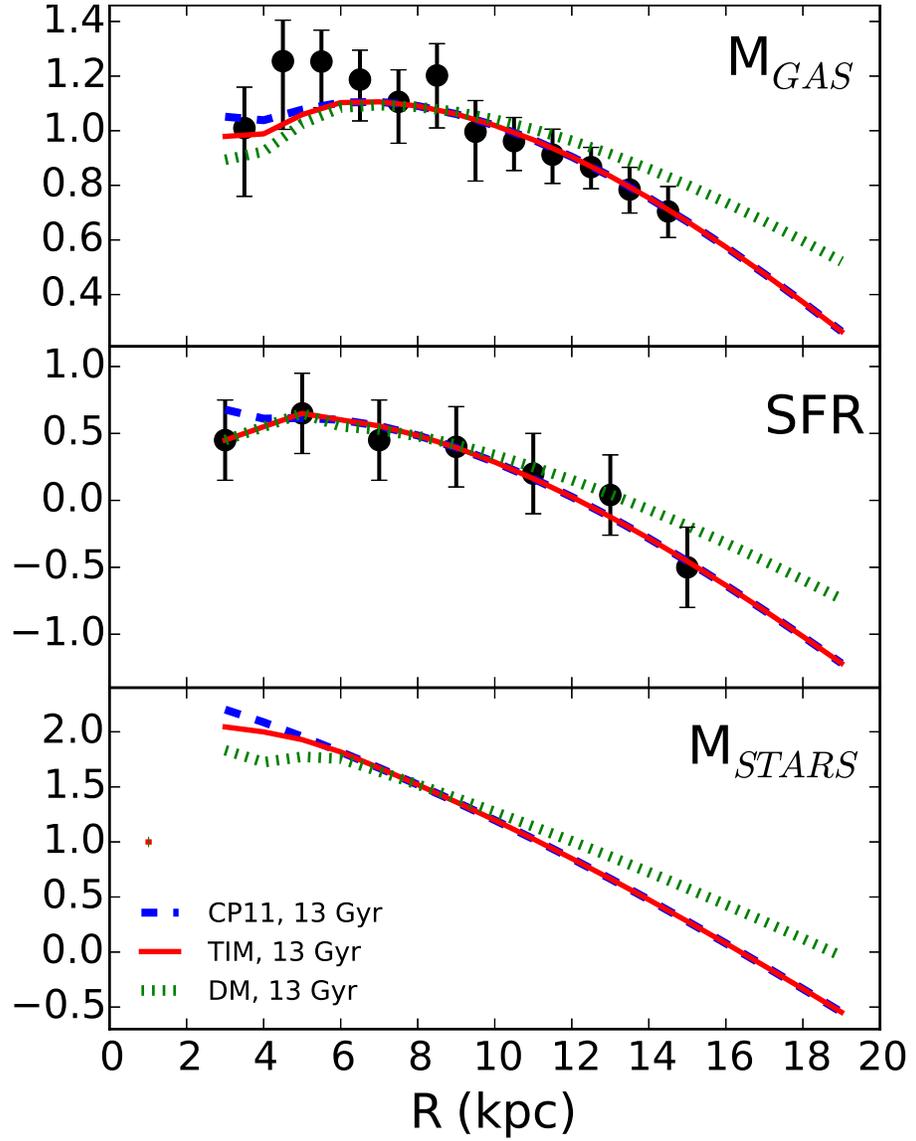}
\end{center}
\caption[f9.eps]{
Current radial distribution  of $M_{gas}$ (\msun pc$^{-2}$),  $SFR$ (\msun pc$^{-2 }$Gyr$^{-1}$), 
and $M_{star}$(\msun pc$^{-2 }$) from CP11, TIM and DM models. Vertical axis in logarithmic 
scale. Observational data by \citet{kennicutt2012}.
\label{Apendix2}}
\end{figure}

\clearpage

\begin{figure}
\begin{center}
\includegraphics[angle=0,scale=0.825]{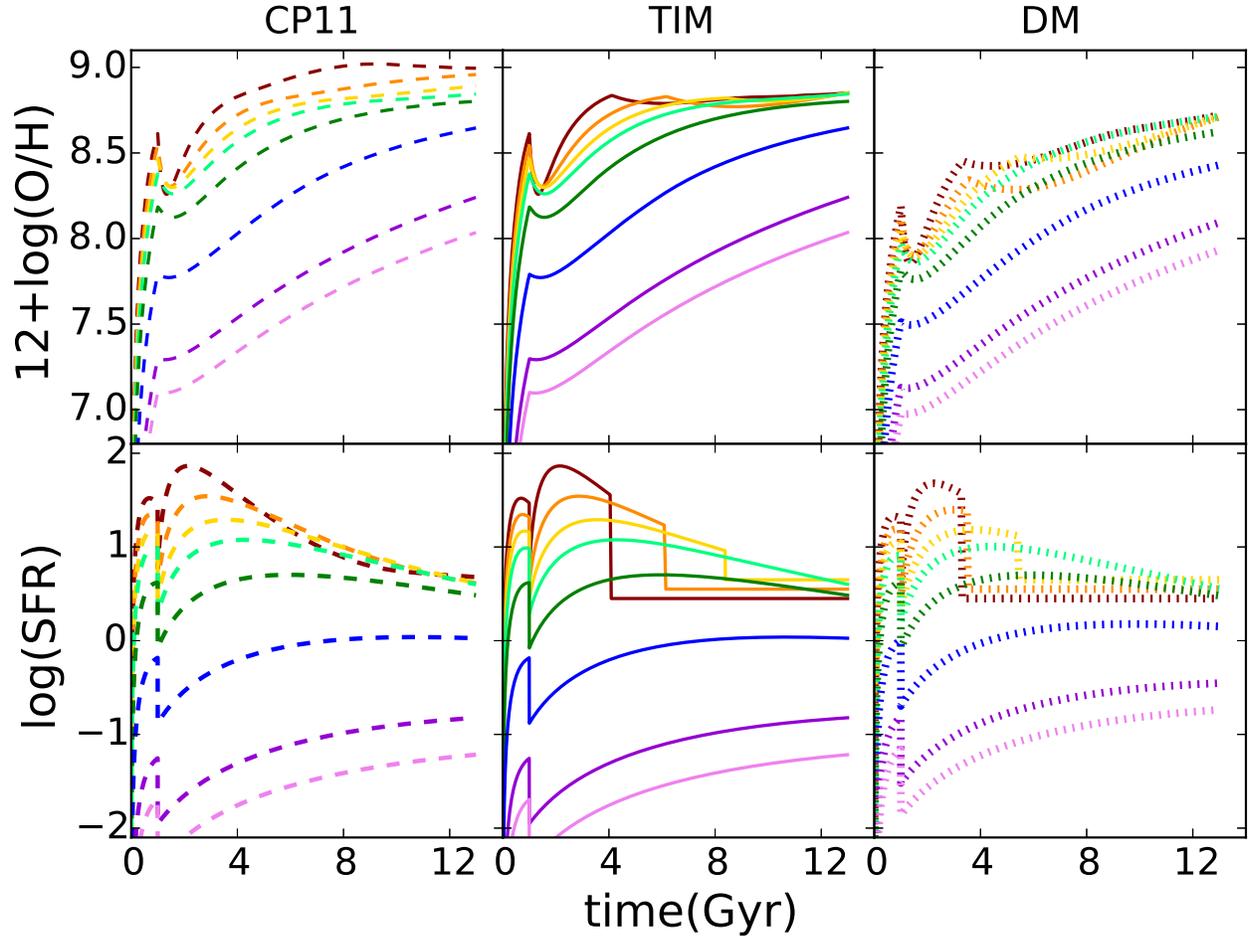}
\end{center}
\caption[f11.eps]{
Evolution of O/H and star formation rate (\msun pc$^{-2 }$Gyr$^{-1}$) inferred from CP11, 
TIM and DM models. Color lines represent different galactocentric radius, as Figure 5.
\label{Apendix4}}
\end{figure}


\begin{thebibliography}{}

\bibitem[Andrievsky et~al.(2016)]{andrievsky2016}
Andrievsky, S. M., Martin, R. P., Kovtyukh, V. V., Korotin, S. A., \& L\'epine, J. R. D. 2016, \mnras, 461, 4256

\bibitem[Asplund et~al.(2009)]{asplund2009}
Asplund, M., Grevesse, N., Sauval, A. J.,  \& Scott, P. 2009, \araa, 47, 481


\bibitem[Belfiore et~al.(2017)]{belfiore2017}
Belfiore, F.,  Maiolino, R.,  Maraston, C. et al. 2017, \mnras, 466, 2570

\bibitem[Berg et~al.(2016)]{berg2016}
Berg, D. A., Skillman, E. D., Henry.  R. B. C., Erb, D. K.,  \& Carigi, L. 2016
 \apj, 827, 126
 
  \bibitem[Bresolin et~al.(2009)]{bresolin2009}
Bresolin, F., Gieren, W., Kudritzki, R.-P.,   et al. 2019,
 \apj, 700, 309
 
 \bibitem[Bresolin et~al.(2016)]{bresolin2016}
Bresolin, F., Kudritzki, R.-P., Urbaneja, M. A., et al. 2016,
 \apj, 830, 64

\bibitem[Carigi \& Hernandez(2008)]{carigi2008}
Carigi, L. \& Hernandez, X., 2008, MNRAS, 390, 582

\bibitem[Carigi \& Peimbert(2008)]{carigipeim2008}
Carigi, L. \& Peimbert, M., 2008, RMxAA., 44, 341

\bibitem[Carigi \& Peimbert(2011)]{carigi2011}
Carigi, L. \& Peimbert, M., 2011,  RMxAA, 47, 139

\bibitem[Cavichia et~al.(2014)]{cavichia2014}
Cavichia, O.,  Moll\'a, M., Costa, R. D. D., \& W. J. Maciel, W. J., 2014, MNRAS, 437, 3688

\bibitem[Chiappini et~al.(1997)]{chiappini1997}
Chiappini, C., Matteucci, F., \& Gratton R. 1997, ApJ, 477, 765

\bibitem[C\^ot\'e et~al.(2016)]{cote2016}
C\^ot\'e, B.,  Ritter, C., O'Shea, B. W., Herwig, F., Pignatari, M., Jones, S., Fryer, C. L. 2016, \apj,  824, 82 

\bibitem[Esp\'{\i}ritu et~al.(2017)]{espiritu2017}
Esp\'{\i}ritu, J. N., Peimbert, A. Delgado-Inglada, G., \& Ruiz, M. T. 2017, RMxAA, 53, 95

\bibitem[Esteban et~al.(2013)]{esteban2013}
Esteban, C., Carigi, L., Copetti, M. V. F., Garc\'{\i}a-Rojas, J., Mesa-Delgado, A., Casta\~neda, H. O., \& P\'equignot, D. 2013, \mnras, 433, 382

\bibitem[Esteban et~al.(2017)]{esteban2017}
Esteban, C., Fang, X., Garc\'{\i}a-Rojas, J., \& Toribio San Cipriano, L. 2017, \mnras, 471, 987

\bibitem[Esteban et~al.(2005)]{esteban2005}
Esteban, C., Garc\'{\i}a-Rojas, J., Peimbert,  M., Peimbert,  A.,
Ruiz, M. T., Rodr\'{\i}guez,  M., \& Carigi,  L., 2005, \apj, 618, L95

\bibitem[Esteban \& Garc\'{\i}a-Rojas (2018)]{esteban2018}
Esteban, C., \& Garc\'{\i}a-Rojas, J., 2018, \mnras, 478, 2315
 
\bibitem[Esteban et~al.(2016)]{esteban2016}
Esteban, C., Mesa-Delgado, A., Morisset, C., \& Garc\'{\i}a-Rojas, J. 2016, \mnras, 460, 2038

\bibitem[Esteban et~al.(2004)]{esteban2004}
Esteban, C., Peimbert, M., Garc\'{\i}a-Rojas, J., Ruiz, M. T., Peimbert, A., \& Rodr\'{\i}guez, M. 2004 \mnras, 355, 229

\bibitem[Fern\'andez-Mart\'{\i}n et~al.(2017)]{fernandez2017}
Fern\'andez-Mart\'{\i}n, A., P\'erez-Montero, E., V\'{\i}lchez, J. M., \& Mampaso, A. 2017, \aap, 597, A84

\bibitem[Fenner \& Gibson (2003)]{fenner2003}
Fenner, Y., \& Gibson, B. K. 2003, PASA, 20, 189

\bibitem[Garc\'{\i}a-Rojas et~al.(2005)]{garcia2005}
Garc\'{\i}a-Rojas, J., Esteban, C.,  Peimbert, A., Peimbert, M., Rodr\'{\i}guez, M., \& Ruiz, M. T. 2005, \mnras, 362, 301

\bibitem[Garc\'{\i}a-Rojas et~al.(2007)]{garcia2007}
Garc\'{\i}a-Rojas, J.,  Esteban, C., Peimbert, A., Rodr\'{\i}guez, M.,  Ruiz, M. T. Peimbert, M., \& Ruiz, M. T. 2007, RMxAA, 43, 3

\bibitem[Garc\'{\i}a-Rojas et~al.(2006)]{garcia2006}
Garc\'{\i}a-Rojas, J.,  Esteban, C.,  Peimbert, M.,  Costado, M. T., Rodr\'iguez, M.,  Peimbert, A., \& Ruiz,  M.T., 2006,  \mnras, 368, 253

\bibitem[Garc\'{\i}a-Rojas et~al.(2004)]{garcia2004}
Garc\'{\i}a-Rojas, J., Esteban, C., Peimbert, M., Rodr\'{\i}guez, M., Ruiz, M. T., \&  Peimbert, A. 2004,  \apjs, 153, 501

\bibitem[Garc\'{\i}a-Rojas et~al.(2014)]{garcia2014}
Garc\'{\i}a-Rojas, J., Sim\'on-D\'{\i}az, S., \& Esteban, C. 2014, \aap, 571, A93

\bibitem[Greggio \& Renzini (1983)]{greggio1983}
Greggio, L. \& Renzini, A. 1983, A\&A, 118, 217

\bibitem[Hayden et~al.(2015)]{hayden2015} 
Hayden, M. R., Bovy, J., Holtzman, J. A. et al. 2015, ApJ, 808, 132

\bibitem[Hern\'andez-Mart\'inez et~al.(2011)]{hernandez2011} 
Hern\'andez-Mart\'inez, L., Carigi, L., Pe\~na, M., Peimbert, M. 2011, A\&A, 535, 118
 
\bibitem[Hirschi (2007)]{hirschi2007} 
Hirschi, R. 2007, \aap, 461, 571 

\bibitem[Kennicutt (1998)]{kennicutt1998}
Kennicutt, R. C.  1998, ApJ, 498, 541

\bibitem[Kennicutt \& Evans(2012)]{kennicutt2012}
Kennicutt, R. C. \& Evans, N. J. 2012, ARA\&A, 50, 531

\bibitem[Korotin et~al.(2014)]{korotin2014}
Korotin, S. A., Andrievsky, S. M., Luck. R. E., L\'epine, J. R. D., Maciel, W. J., \& Kovtyukh, V. V. 2014, \mnras, 444, 3301

\bibitem[Kroupa et~al. (1993)]{kroupa1993} 
Kroupa, P., Tout, C. A., \& Gilmore, G. 1993, MNRAS, 262, 545  

\bibitem[Kubryk et~al.(2015a)]{kubryc2015a}
Kubryk, M., Prantzos, N.,  \&  Athanassoula, E. 2015a, A\&A, 580, A126

\bibitem[Kubryk et~al.(2015b)]{kubryc2015b}
Kubryk, M., Prantzos, N.,  \&  Athanassoula, E. 2015b, A\&A, 580, A127

\bibitem[Lian et~al. (2017)]{lian2017} 
Lian, J.,  Yan, R., Blanton, M., \& Kong, X. 2017
\mnras, 472, 4679

\bibitem[Luck et~al. (2013)]{luck2013} 
Luck, R. E., Andrievsky S. M., Korotin S. N., \& Kovtyukh V. V., 2013,  AJ, 146, 18

\bibitem[Luck \& Lambert(2011)]{luck2011}
Luck, R. E. \& Lambert, D. L. 2011, \aj, 142, 137

\bibitem[Maeder (1992)]{maeder1992}
Maeder A., 1992, A\&A, 264, 105

\bibitem[Marigo et~al. (1996)]{marigo1996} 
Marigo, P., Bressan, A., \& Chiosi, C., 1996, A\&A, 313, 545

\bibitem[Marigo et~al. (1998)]{marigo1998} 
Marigo, P., Bressan, A., \& Chiosi, C., 1998, A\&A, 331, 564

\bibitem[Martin et~al.(2015)]{martin2015}
Martin, R. P., Andrievsky, S. M., Kovtyukh, V. V., Korotin, S. A., Yegorova, I. A., \& Saviane, I. 2015, \mnras, 449, 4071

\bibitem[Mart\'{\i}nez-Medina et~al.(2017)]{martinez2017}
Mart\'{\i}nez-Medina, L. A., Pichardo, B., Peimbert, A., \& Carigi, L. 2017, MNRAS, 468, 3615

\bibitem[Matteucci \& Chiappini (1999)]{matteucci1999}
Matteucci, F., \& Chiappini, C. 1999, in Chemical Evolution from Zero to High Redshift, ed. J. R. Walsh \& M. R. Rosa (Berlin: Springer-Verlag), 83

\bibitem[Meynet \& Maeder (2002)]{meynet2002} 
Meynet, G., \&  Maeder, A. 2002,  \aap, 390, 561

\bibitem[Minchev et~al.(2013)]{minchev2013}
Minchev, I., Chiappini, C. \& Martig, M., 2013, A\&A 558, A9

\bibitem[Minchev et~al.(2014)]{minchev2014}
Minchev, I., Chiappini, C. \& Martig, M., 2014, A\&A 572, A92 

\bibitem[Mishurov \& Tkachenko(2018)]{mishurov2018}
Mishurov, Yu. N. \& Tkachenko, V. 2018, \mnras, 473, 3700

\bibitem[Moll\'a et~al.(2015)]{molla2015}
Moll\'a, M., Cavichia, O.,  Gavil\'an, M., and Gibson, B. K. 2015, 
\mnras, 451, 3693

\bibitem[Moll\'a et~al.(2016)]{molla2016}
Moll\'a, M., D\'iaz, A. I., Gibson, B. K., Cavichia, O.,
L\'opez-S\'anchez, A. R. 2016
\mnras, 462, 1329

\bibitem[Nieva \&  Sim\'on-D\'{\i}az(2011)]{nieva2011}
Nieva, M.-F. \&  Sim\'on-D\'{\i}az, S. 2011, \aap, 532,  2

\bibitem[Nieva \& Przybilla (2012)]{nieva2012}
Nieva, M.-F. \& Przybilla, N. 2012, A\&A, 539, A143

\bibitem[Nomoto et~al.(1997)]{nomoto1997}
Nomoto, K., Iwamoto, K., Nakasato, N., Thielemann, F.-K., Brachwitz, F.,
et al. 1997, Nucl. Phys. A., 621, 467

\bibitem[Nuza et~al.(2018)]{nuza2018}
Nuza, S. E., Scannapieco, C., Chiappini, C., Junqueira, T. C.,
Minchev, I., \& Martig, M. 2018
\mnras, submitted (arxiv:1805.06428)

\bibitem[Peimbert \& Peimbert(2010)]{peimbert2010}
Peimbert, A. \& Peimbert, M. 2010, \apj, 724, 791

\bibitem[Peimbert(1967)]{peimbert1967}
Peimbert, M., 1967, \apj, 151, 825

\bibitem[Peimbert \& Costero(1969)]{peimbert1969}
Peimbert, M. \& Costero, R. 1969, BOTT, 5, 3

\bibitem[Peimbert et~al.(2017)]{peimbert2017}
Peimbert, M., Peimbert, A., \& Delgado-Inglada, G.  2017, \pasp, 129,082001

\bibitem[Pe\~na-Guerrero et~al.(2012)]{pena2012}
Pe\~na-Guerrero, M. A., Peimbert, A., \& Peimbert, M. 2012, \apj, 756, L14

\bibitem[P\'erez-Montero(2017)]{perez2017}
P\'erez-Montero, E. 2017, \pasp, 129, 03001

\bibitem[Portinari et~al.(1998)]{portinari1998}
Portinari, L., Chiosi, C., \& Bressan, A., 1998, A\&A, 334, 505

\bibitem[Portegies Zwart(2009)]{portegies2009}
Portegies Zwart S. F., 2009, ApJ, 696, L13

\bibitem[Prantzos(2016)]{prantzos2016}
Prantzos, N. 2016, Astronomische Nachrichten, 337, 953

\bibitem[Renda et~al.(2005)]{renda2005}
Renda, A., Kawata, D., Fenner, Y. and Gibson. B. K. 2005, MNRAS, 356, 1071

\bibitem[Robles-Valdez et~al.(2013)]{robles2013}
Robles-Valdez, F., Carigi, L., Peimbert, M. 2013, MNRAS, 429, 2351

\bibitem[Rolleston et~al.(2000)]{rolleston2000}
Rolleston, W. R. J., Smartt, S. J., Dufton, P. L., \& Ryans, R. S. I. 2000, \aap, 363, 537

\bibitem[Rudolph et~al.(2006)]{rudolph2006}
Rudolph, A. L., Fich, M., Bell, G. R., et al. 2006, \apjs, 162, 346

\bibitem[Romano et~al.(2005)]{romano2005}
Romano, D., Chiappini, C., Matteucci, F., and Tosi, M. 2005, A\&A, 430, 491

\bibitem[Romano et~al.(2010)]{romano2010}
Romano, D., Karakas, A. I.,  Tosi, M., and Matteucci, F. 2010,  A\&A,
522, A32

\bibitem[S\'anchez-Menguiano et al.(2018)]{sanchez2018}
S\'anchez-Menguiano, L., S\'anchez, S. F., P\'erez, I., Ruiz-Lara, T, Galbany, L., et al. 2018, A\&A, 609, 119

\bibitem[Shi et~al.(2018)]{shi2018}
Shi, Y., Yan, L., Armus, L., Gu, Q., Helou, G., Qiu, K.,
Gwyn, S., Stierwalt, S., Fang, M., Chen, Y. et al., 2018, \apj, 853, 149

\bibitem[Smartt et al.(2001)]{smartt2001}
Smartt, S. J., Venn, K. A., Dufton, P. L.,  et al.,  2001, \aap, 367, 86

\bibitem[Toribio San Cipriano et al.(2016)]{toribio2016}
Toribio San Cipriano, L., Garc\'{\i}a-Rojas, J., Esteban, C.,  Bresolin, F., \& Peimbert, M.,
2016, \mnras, 458, 1866

\bibitem[Wielen, Fuchs, \& Dettbarn(1996)]{wielen1996}
Wielen, R., Fuchs, B., \& Dettbarn, C., 1996, \aap, 314, 438

\bibitem[Woosley \& Weaver(1995)]{woosley1995}
Woosley, S. E. \&  Weaver, T. A., 1995, ApJS, 101, 181

\bibitem [Yin et al.(2009)]{yin2009}
Yin, J., Hou, J. L., Prantzos, N., Boissier, S., Chang, R. X., Shen, S. Y., \&  Zhang, B., 2009, A\&A, 505, 497

\end{thebibliography}
\end{document}